\documentclass[a4paper,11pt]{article}
\pdfoutput=1

\usepackage{jheppub}
\usepackage[T1]{fontenc}

\usepackage{amsmath,amsfonts,amssymb}
\usepackage{booktabs}
\usepackage{cancel}
\usepackage{caption}
\usepackage{graphicx}
\usepackage{multirow}
\usepackage{pifont}
\usepackage{siunitx}
\usepackage{slashed}
\usepackage{subcaption}

\newcommand\Y{\ding{51}}
\newcommand\N{\ding{55}}

\newcommand\Lmass{\mathcal{L}_{\text{mass}}}
\newcommand\Llinear{\mathcal{L}^{\text{lin}}}
\newcommand\Lquad{\mathcal{L}^{\text{quad}}}

\newcommand\Lcal{\mathcal{L}}
\newcommand\Ocal{\mathcal{O}}
\newcommand\coef[1]{#1 \,}

\newcommand\Otphi{\Ocal_{t \phi}}
\newcommand\Obphi{\Ocal_{b \phi}}
\newcommand\Ophit{\Ocal_{\phi{} t}}
\newcommand\Ophib{\Ocal_{\phi{} b}}
\newcommand\Ophitb{\Ocal_{\phi{} tb}}
\newcommand\Ophiq[1]{\Ocal_{\phi{} q}^{(#1)}}
\newcommand\OtB{\Ocal_{t B}}
\newcommand\ObB{\Ocal_{b B}}
\newcommand\OtW{\Ocal_{t W}}
\newcommand\ObW{\Ocal_{b W}}
\newcommand\OtG{\Ocal_{t G}}
\newcommand\ObG{\Ocal_{b G}}
\newcommand\OphiG{\Ocal_{\phi G}}

\newcommand\sigmaProd{\sigma_{pp \to{} Q\bar{Q}}}

\newcommand\LambdaQCD{\Lambda_{\text{QCD}}}
\newcommand\LambdaD{\Lambda_{\text{disp}}}
\newcommand\LambdaLL{\Lambda_{\text{long lived}}}

\newcommand\hc{\text{h.c.}}
\newcommand\re{\operatorname{Re}}

\newcommand\DLR{\overset{\leftrightarrow}{D}}

\newcommand\multiplet[1]{\left(\begin{array}{c} #1 \end{array}\right)}

\newcommand\SM{\text{SM}}

\renewcommand\mod[1]{\; (\text{mod} \, #1)}

\title{Vector-like quarks with non-renormalizable interactions}

\author[a]{J. C. Criado,}
\author[a]{M. P\'erez-Victoria}

\affiliation[a]{CAFPE and Departamento de F\'{\i}sica Te\'orica y del Cosmos, \\ Universidad de Granada, Campus de Fuentenueva, E-18071 Granada, Spain}

\emailAdd{jccriadoalamo@ugr.es}
\emailAdd{mpv@ugr.es}

\abstract{We study the impact of the leading non-renormalizable terms in the effective field theory that describes general extensions of the Standard Model with vector-like quarks that can decay into Standard Model particles. Dropping the usual assumption of renormalizability has several phenomenological consequences for the production and decay of the heavy quarks and also for Higgs physics. The most dramatic effects, including those associated with a long lifetime, occur for vector-like quarks with non-standard quantum numbers.
}

\begin{document}

\maketitle
\flushbottom


\section{Introduction}

The fermions in the Standard Model (SM) can be classified as leptons and quarks, according to their respective transformations as singlets or triplets under the colour gauge group.
Additional spin 1/2 particles with these colour quantum numbers are often
considered by theorists and experimentalists in the quest for new physics. This
is motivated both by their rich phenomenology and by their frequent occurrence
in explicit models. Due to constraints from electroweak precision data and Higgs
physics, these new particles cannot acquire their mass only from the Higgs
vacuum expectation value (vev). The necessary gauge-invariant mass term requires left-handed
and right-handed components, transforming in the same way, not only under the
colour group, but also, unlike the SM fermions, under the electroweak gauge
group. Dirac fermions with this property are known as vector-like
fermions.\footnote{A gauge-invariant mass is also possible for Majorana fermions in real representations of the SM gauge group, which can also be considered vector-like fermions.} Two important properties of
vector-like fermions is that all their observable effects decouple when their
mass is taken to infinity and that they never give rise to anomalies of the SM
gauge group. In this paper we concentrate on vector-like extra quarks. The
indirect and direct effects of general vector-like leptons have been analyzed in
refs.~\cite{delAguila:2008pw} and~\cite{delAguila:2008cj}, respectively.

Vector-like quarks appear in many motivated extensions of the SM, for diverse reasons. In models with additional symmetries, they may complete multiplets that include SM fermions \cite{Fritzsch:1975sr,Kearney:2013cca,Agashe:2004rs}. They may also be necessary for the cancellation of the anomalies of an extended gauge group~\cite{Batra:2005rh}. In models with (partially) composite quarks~\cite{Kaplan:1991dc}, they emerge effectively as resonances, while in models in extra dimensions, they show up as Kaluza-Klein modes when the quarks propagate through the bulk~\cite{Gherghetta:2000qt}. Vector-like quarks are also used to relax the bounds from precision observables~\cite{Choudhury:2001hs} or to avoid strong fine tuning in the Higgs sector~\cite{ArkaniHamed:2001nc,Contino:2006qr}. Here, we will not worry about the origin of the vector-like quarks or the details of the model in which they appear. Instead, we follow a systematic model-independent approach by studying a general effective field theory that describes the new quarks and their interactions with the SM fields. Our conclusions can be easily translated to specific models. 

Most analyses of vector-like quarks so far have assumed renormalizable interactions (we comment on exceptions below). At the renormalizable level, the 
possible gauge-invariant interactions of the extra quarks in the electroweak symmetric phase are the ones with the gauge bosons, determined by their quantum numbers, and Yukawa interactions involving  either two extra quarks or one extra quark and one SM quark.  Upon electroweak breaking, the Yukawa couplings give rise to off-diagonal terms in the quark mass matrix, which  translate into the mixing of mass eigenstates in the interaction terms with the $Z$ and $W$  bosons and the Higgs (beyond the mixing in the original Yukawas).  Many of the observable effects of the new quarks, such as their decay into SM particles, their single production and the induced modifications of the light-quark couplings, are associated to their mixing with the SM quarks, which is suppressed when their gauge-invariant mass is larger than the $Z$ mass~\cite{delAguila:1982fs}. This suppression is stronger for heavy vector-like quarks that are not directly connected by Yukawa couplings to the SM quarks. Therefore, the effects of mixing are sizable only in the presence of vector-like quarks with gauge quantum numbers that allow for such couplings. Assuming that electroweak breaking is mostly triggered by the vev $v$ of one or more Higgs doublets, in agreement with limits on the $\rho$ parameter, there are seven different multiplets of vector-like quarks that carry the appropriate quantum numbers. They are shown in the first seven rows of table~\ref{tab:dim5-reps}. Note that these are the only vector-like quarks that can couple linearly to SM operators in a renormalizable theory.\footnote{More generally, the only extra fermion fields of spin 1/2 that can couple linearly to SM operators in a renormalizable theory are either colour singlets or colour triplets, that is, either leptons or quarks.} Vector-like quarks with this property will be called ``renormalizable'' vector-like quarks (RVLQ), even if they can also have non-renormalizable interactions. Their components have electric charges in the set $\{\pm 1/3, \pm 2/3, \pm 4/3, \pm 5/3\}$. The most general renormalizable extension of the SM with arbitrary combinations of the seven types of RVLQ was explicitly written in ref.~\cite{delAguila:2000rc}.  In that work, the leading indirect effects beyond the SM, including flavour-changing neutral currents, right-handed charged currents and a non-unitary CKM matrix, were studied by integrating the heavy quarks out and using the results in ref.~\cite{delAguila:2000aa} for the relevant flavourful part of the SM effective field theory (SMEFT) at dimension six. The loop contributions of these multiplets to oblique parameters have also been calculated in refs.~\cite{Lavoura:1992np,Carena:2006bn,Anastasiou:2009rv}.  Regarding direct searches, refs.~\cite{AguilarSaavedra:2009es,Aguilar-Saavedra:2013qpa} provide a comprehensive and detailed guide to the LHC phenomenology of minimal renormalizable extensions of the SM with vector-like quarks that mix dominantly with the third family. Several other works have been devoted to collider searches of RVLQ, see for instance refs.~\cite{Cacciapaglia:2010vn,Beauceron:2014ila,Barducci:2014ila,Moretti:2016gkr,Carvalho:2018jkq}.

In the present work, we extend this framework by allowing non-renormalizable interactions. This allows us to assess the robustness of the standard limits on vector-like quarks and to explore possible new observable signals. We consider an effective Lagrangian, invariant under the SM gauge symmetry and constructed with the SM fields (including the Higgs doublet) and the vector-like quarks. For simplicity, we will consider simple extensions with only one quark multiplet at a time. 
The cutoff scale $\Lambda$ of the effective Lagrangian is required to be larger than all the mass scales in the theory, and in particular larger than the gauge invariant mass $M$ of the new quark. All the possible particles not included in the effective Lagrangian, such as additional extra quarks or extra scalars, are assumed to be heavier than $\Lambda$; their effects are then encoded into the Wilson coefficients of the effective theory. As usual, the effective Lagrangian is to be expanded in inverse powers of $\Lambda$. When $\Lambda$ is much higher than the probed energies $E$ and the Higgs vev $v$, all the effects of higher-dimensional operators will be suppressed by powers of $E/\Lambda$ and/or $v/\Lambda$ with respect to the effects of the renormalizable ones and will typically give rise to small corrections to the known results. However, some processes may require the presence of higher-dimensional interactions, which will then provide the leading contributions. In particular, this is always the case for quark multiplets that can only couple linearly and gauge invariantly to the SM fermions at the non-renormalizable level. As we will see, the phenomenology of these multiplets can indeed be different from the one of RVLQ.

In fact, relaxing the requirement of renormalizability enlarges the list of vector-like quarks that can mix with the SM ones and, more generally, have linear couplings with SM operators.\footnote{Interestingly, non-renormalizable linear interactions of other colour representations, beyond singlets and triplets, are also allowed.} The number of different multiplets with this property is finite at each order in $1/\Lambda$ and increases with the order in this expansion. In this paper, we only study explicitly the leading corrections to renormalizable theories with vector-like quarks. So, we will truncate the effective Lagrangian at order $1/\Lambda$, that is, we will consider only operators of canonical dimension $n\leq 5$. The quark multiplets that can have linear couplings to this order are collected in table~\ref{tab:dim5-reps}. As can be checked there, there are five new multiplets, in addition to the seven RVLQ. The new ones will be called "non-renormalizable" vector-like quarks (NRVLQ). The only gauge-invariant operator that can be built with the SM fields at dimension 5 is the Weinberg operator, which involves only leptons and can thus be ignored in our context. The rest of dimension-5 operators always contain at least one of the extra quarks in table~\ref{tab:dim5-reps}.  In order to simplify the analysis, we will assume that the extra quarks do not couple to the first two SM families. This assumption can easily be dropped, at the price of introducing more free parameters. We study the mixing with the third family of SM quarks and the associated phenomenology, including indirect effects on electroweak and Higgs observables and the production and decay of the new quarks.  We will see that for some multiplets there are new single production mechanisms and new decay channels, which can be sizable in some regions of parameter space. A significant feature of the vector-like quarks without renormalizable interactions is that their widths are suppressed. For dimensionless couplings of order 1 and a cutoff $\Lambda$ larger than 5 TeV, it turns out that their lifetimes are larger than the typical QCD times and thus non-perturbative effects, including hadronization, will take place before decay. For still larger values of $\Lambda$, the NRVLQ, or more precisely the hadrons they form, will be long lived. These quarks would then elude the usual searches, which assume prompt decays, and lead instead to alternative signatures, such as tracks with anomalous ionization, long time of flight or displaced vertices.

Extra quarks with non-renormalizable interactions have been studied before in the context of pseudo-Goldstone composite Higgs models~\cite{Contino:2006qr,Matsedonskyi:2012ym,DeSimone:2012fs,Matsedonskyi:2015dns}. This is a particular subclass of the theories included in our general model-independent framework, with $\Lambda$ identified with the symmetry breaking scale $f$. But in the pseudo-Goldstone scenario, the assumed symmetry breaking pattern allows to easily resum the $1/f$ expansion. Then, $f$ can be pretty low without loosing predictive power.\footnote{The effective descriptions of these models are valid up to a cutoff higher than $f$, associated to additional resonances or strong coupling. In explicit holographic models, these effects are incorporated and the cutoff can be much higher for many purposes~\cite{Rattazzi:2003ea}.} The vector-like quarks in those models belong to multiplets of an extended symmetry and, for the popular choices in the literature, decompose under the SM gauge group into a subset of the seven RVLQ representations. Here, we want to follow a model-independent approach, so we do not make any assumptions about the nature of the Higgs, about symmetries beyond the SM ones or about the representations of the quarks (except for the requirement of linear interactions). Another study of non-renormalizable interactions for new quarks, similar in spirit to the one in this paper, was presented in ref.~\cite{Fajfer:2013wca}. There, the first three multiplets in table~\ref{tab:dim5-reps}, coupled via operators involving the Higgs, were considered. We generalize this work by including all the relevant multiplets and operators at dimension 5. In particular, we consider multiplets without dimension-4 interactions, which present the most dramatic changes with respect to the usual phenomenology of vector-like quarks. On the other hand, the flavour structure we assume is more restrictive than the one in ref.~\cite{Fajfer:2013wca}, which allowed for couplings to the light families of SM quarks. 

We have implemented in {\sc FeynRules 2.0} ~\cite{Alloul:2013bka} the effective theory for each vector-like multiplet in table~\ref{tab:dim5-reps}. All the simulations have been performed with {\sc MadGraph5\_aMC@ NLO}~\cite{Alwall:2011uj,Alwall:2014bza} with the UFO files generated with {\sc FeynRules}. 

The paper is organized as follows. In section~\ref{sec:extensions}, we introduce the effective theory for vector-like quarks, find the constraints on quantum numbers for linear interactions and write explicitly the general Lagrangian for an arbitrary multiplet with all the operators of dimension up to 5. We also comment briefly on the possible ultraviolet (UV) origin of the non-renormalizable operators. In section~\ref{sec:mixing}, we diagonalize the mass matrices that appear in the Higgs phase for the components with the same electric charges as the SM quarks. Section~\ref{sec:indirect} is devoted to indirect effects of the new quarks and to the corresponding limits from Higgs, electroweak and top data. Production at hadron colliders is discussed in section~\ref{sec:production}, while the decay of the new quarks is examined in section~\ref{sec:decay}. We present our conclusions in~\ref{sec:conclusions}. Three appendices are devoted to some important technical results that are used in the main text. In appendix~\ref{app:sm-operators}, we obtain a necessary condition for the linear coupling of new fields to SM operators of arbitrary dimension. In appendix~\ref{app:mass-limits}, we find a simple formula to reinterpret the mass limits provided by the LHC collaborations in the case with additional decay modes.  Our method is based on the one in ref.~\cite{Aguilar-Saavedra:2017giu}. Finally, in appendix~\ref{app:diagonal} we explain why the branching ratios to Higgs and Z bosons are approximately equal for all multiplets but one.

\section{Non-renormalizable extensions of the Standard Model with vector-like quarks}

\label{sec:extensions}

Let us consider a general local effective Lagrangian $\Lcal$ describing extensions of the SM with extra vector-like
quarks. The effective theory will be valid for energies smaller than a certain cutoff $\Lambda$, which is larger than all the mass scales in the theory, including the mass $M$ of the extra quarks. $\Lcal$ must be invariant under Lorentz and $SU(3)\times SU(2) \times U(1)$ gauge transformations. We assume that the latter is linearly realized. The degrees of freedom that appear in the Lagrangian are the SM fields, including the Higgs doublet, and the new spinors, which transform as triplets under $SU(3)$. We do not impose renormalizability. Up to this point, the $SU(2) \times U(1)$ representation of the new quarks is completely arbitrary, except for the vector-like condition, which requires that the new quarks be grouped in pairs of Weyl spinors belonging to the same representation. Now, let us introduce our main non-trivial restriction: the new quarks are assumed to have linear couplings to the SM fields. In other words, there exist interaction terms involving products of SM fields and a single power of the extra quark field. This assumption is motivated by phenomenology, as explained in the introduction, and has the crucial advantage of restricting the irreducible representations of $SU(2)\times U(1)$ to a finite set, when the $1/\Lambda$ expansion of $\Lcal$ is truncated at any order. Indeed, the isospin and hypercharge of a quark coupling linearly to an SM operator $\mathcal{O}$ is given directly by the isospin and hypercharge of $\mathcal{O}$, which belong to a finite set for a fixed dimension of the operator. 

Moreover, we find in appendix~\ref{app:sm-operators} a general constraint over the representation of
any Standard Model operator, and thus of any field with a gauge-invariant linear coupling.\footnote{This condition has been given before, in a different form, in ref.~\cite{Kats:2012ym}.}  In the case of colour triplets, it reads
\begin{equation}
  T + Y + 1/3 \in \mathbb{Z},
  \label{eq:quarks-reps}
\end{equation}
with $T$ the isospin of the $SU(2)$ representation and $Y$ the hypercharge. It is also true that, given a representation of $SU(2) \times U(1)$ satisfying
eq.~\eqref{eq:quarks-reps}, there is a product of Standard Model fields that
produces this representation. Indeed, consider first the products $\phi^k {(\phi^*)}^l$
of the Higgs doublet and its conjugate. They generate all representations with
$T + Y \in \mathbb{Z}$. Then, the operators of the form
$\phi^k {(\phi^*)}^l q$ give all the possibilities satisfying
eq.~\eqref{eq:quarks-reps}. So this formula allows to find easily the quark multiplets with linear couplings. 

Higher-dimensional multiplets couple linearly to the Standard Model through
higher-dimensional operators. Therefore, the effects of higher-dimensional
multiplets tend to be more suppressed than the lower-dimensional ones. As we have just explained, at each
order in inverse powers of the cutoff $\Lambda$, which is given by the dimension of the operators, there is a finite number of
multiplets with linear couplings to SM fields. This number increases with the order in $1/\Lambda$. We focus in the following on the
next-to-leading order in this expansion, which is $O(1/\Lambda)$. Equivalently,
we impose a maximum dimension of 5 for the operators in the effective
Lagrangian. There are twelve possible multiplets with linear couplings at this
order, listed in table~\ref{tab:dim5-reps}. The ones in the first seven rows, called RVLQ in this paper, can have linear
interactions of dimension 4. These are the multiplets that have been studied in the past. For natural values of the couplings, the dimension-5 operators will generate small
corrections to their properties. The remaining five multiplets, which we call NRVLQ, cannot have dimension-4 linear couplings. Therefore, for these multiplets the dimension-5
interactions will give leading-order effects. Let us stress that RVLQ can have non-renormalizable linear interactions and that NRVLQ have renormalizable quadratic interactions with the gauge fields, besides the kinetic and mass terms. 
Let us also note in passing that, besides the singlet and triplet representations, other irreducible representations of
$SU(3)$ are possible for spin-1/2
particles with dimension-5 linear couplings to the Standard Model. The extra eight
possibilities for their representations under
$SU(3) \times SU(2) \times U(1)$ are:
\begin{gather}
  {(6, 1)}_{-2/3}, \, {(6, 1)}_{1/3}, \, {(6, 2)}_{-1/6}, \,
  {(8, 1)}_1, \, {(8, 2)}_{1/2}, \,
  {(15, 1)}_{2/3}, \, {(15, 1)}_{-1/3}, \, {(15, 2)}_{1/6}.
\end{gather}

Coming back to extra quarks, the dimension-5 operators containing exactly one vector-like quark can have one of
the following two schematic forms: $\bar{Q}\phi\phi q$ and
$\bar{Q}\sigma^{\mu\nu} q F_{\mu\nu}$, where $\phi$ is the Higgs doublet, $q$ and $Q$ represent SM and extra quark multiplets, respectively, and $F_{\mu\nu}$ is the field-strength tensor of a SM gauge field. We do not consider operators with the
field content $\bar{Q}\phi q D$, with $D$ a covariant derivative, because they can be eliminated using integration
by parts and field redefinitions, up to $O(1/\Lambda^2)$ corrections. The interactions allowed for each multiplet
are presented in table~\ref{tab:dim5-reps}. It is important to note that the interactions of the form $\bar{Q}\phi\phi q$ will typically give physical effects suppressed by powers of $v/\Lambda$, while the effects of interactions of the form $\bar{Q}\sigma^{\mu\nu} q F_{\mu\nu}$ are suppressed by powers of $E/\Lambda$, with $E$ the characteristic energy of the process ($E \simeq M$ for on-shell extra quarks). In the rest of this paper, we study
the theories defined by adding each of the possible multiplets at a time. The dimension-5 effective Lagrangian for any such theory with one multiplet
$Q$ is
\( \mathcal{L} = \mathcal{L}_{\text{SM}} + \mathcal{L}^{\text{free}}_{Q} +
(\Llinear_Q + \Lquad_Q + \hc)\), with
\begin{align}
  \mathcal{L}^{\text{free}}_{Q}
  &=
  \bar{Q} (i\slashed{D} - M) Q,
  \\
  -\Llinear_U
  &=
  \coef{\lambda_i} \bar{U}_R \tilde{\phi}^\dagger q_{Li}
  + \coef{y_i} (\bar{U}_L u_{Ri}) (\phi^\dagger \phi)
  + \coef{w_{Bi}} \bar{U}_L \sigma^{\mu\nu} u_{Ri} B_{\mu\nu}
  + \coef{w_{Gi}} \bar{U}_L \lambda^A \sigma^{\mu\nu} u_{Ri} G^A_{\mu\nu},
  \\
  -\Llinear_D
  &=
  \coef{\lambda_i} \bar{D}_R \phi^\dagger q_{Li}
  + \coef{y_i} (\bar{D}_L d_{Ri}) (\phi^\dagger \phi)
  + \coef{w_{Bi}} \bar{D}_L \sigma^{\mu\nu} d_{Ri} B_{\mu\nu}
  + \coef{w_{Gi}} \bar{D}_L \lambda^A \sigma^{\mu\nu} d_{Ri} G^A_{\mu\nu},
  \\
  -\Llinear_{Q_1}
  &=
  \coef{\lambda_{ui}} \bar{Q}_{1L} \tilde{\phi} u_{Ri}
  + \coef{\lambda_{di}} \bar{Q}_{1L} \phi d_{Ri}
  + \coef{y_{ui}} (\bar{Q}_{1R} \tilde{\phi}) (\tilde{\phi}^\dagger q_{Li})
  + \coef{y_{di}} (\bar{Q}_{1R} \phi) (\phi^\dagger q_{Li})
  \nonumber \\
  &\phantom{=}
  + \coef{w_{Bi}} \bar{Q}_{1R} \sigma^{\mu\nu} q_{Li} B_{\mu\nu}
  + \coef{w_{Wi}} \bar{Q}_{1R} \sigma^a \sigma^{\mu\nu} q_{Li} W^a_{\mu\nu}
  + \coef{w_{Bi}} \bar{Q}_{1R} \lambda^A \sigma^{\mu\nu} q_{Li} G^A_{\mu\nu},
  \\
  -\Llinear_{Q_7}
  &=
  \coef{\lambda_i} \bar{Q}_{7L} \phi u_{Ri}
  + \coef{y_i} (\bar{Q}_{7R} \phi) (\tilde{\phi}^\dagger q_{Li}),
\end{align}

\begin{align}
  -\Llinear_{Q_5}
  &=
  \coef{\lambda_i} \bar{Q}_{5L} \tilde{\phi} d_{Ri}
    + \coef{y_i} (\bar{Q}_{5R} \tilde{\phi}) (\phi^\dagger q_{Li}),
  \\
  -\Llinear_{T_1}
  &=
  \coef{\lambda_i} \bar{T}^a_{1R} \phi^\dagger \sigma^a q_{Li}
  + \coef{y_{ui}} \bar{T}^a_{1L} u_{Ri} \phi^\dagger \sigma^a \tilde{\phi}
  + \coef{y_{di}} \bar{T}^a_{1L} d_{Ri} \phi^\dagger \sigma^a \phi
  + \coef{w_i} \bar{T}^a_{1L} \sigma^{\mu\nu} d_{Ri} W^a_{\mu\nu},
  \\
  -\Llinear_{T_2}
  &=
  \coef{\lambda_i} \bar{T}^a_{2R} \tilde{\phi}^\dagger \sigma^a q_{Li}
  + \coef{y_{ui}} \bar{T}^a_{2L} u_{Ri} \phi^\dagger \sigma^a \phi
  + \coef{y_{di}} \bar{T}^a_{2L} d_{Ri} \tilde{\phi}^\dagger \sigma^a \phi
  + \coef{w_i} \bar{T}^a_{2L} \sigma^{\mu\nu} u_{Ri} W^a_{\mu\nu},
  \\
  -\Llinear_{T_4}
  &=
  \coef{y_i} \bar{T}^a_{4L} d_{Ri} \phi^\dagger \sigma^a \tilde{\phi},
  \\
  -\Llinear_{T_5}
  &=
  \coef{y_i} \bar{T}^a_{5L} u_{Ri} \tilde{\phi}^\dagger \sigma^a \phi,
  \\
  -\Llinear_{F_1}
  &=
  \coef{y_i} \bar{F}^a_{1R} C^a_{bc} q_{Lic} \phi^\dagger \sigma^b \phi
  + \coef{w_i} \bar{F}^a_{1R} C^a_{bc} \sigma^{\mu\nu} q_{Lic} W^b_{\mu\nu},
  \\
  -\Llinear_{F_5}
  &=
  \coef{y_i} \bar{F}^a_{5R} C^a_{bc} q_{Lic} \phi^\dagger \sigma^b \tilde{\phi},
  \\
  -\Llinear_{F_7}
  &=
  \coef{y_i} \bar{F}^a_{7R} C^a_{bc} q_{Lic}  \tilde{\phi}^\dagger \sigma^b \phi,
  \\
  - \Lquad_Q
  &=
  \coef{W_B} (\bar{Q}_L \sigma^{\mu\nu} Q_R) B_{\mu\nu}
  + \coef{W_W} (\bar{Q}_L \sigma^{\mu\nu} \mathbf{T}^a_{Q} Q_R) W^a_{\mu\nu}
  + \coef{W_G} (\bar{Q}_L \sigma^{\mu\nu} \mathbf{t}^A_{Q} Q_R) G^A_{\mu\nu}
  \nonumber \\
  &\phantom{=}
  + \coef{Y_1} (\bar{Q}_L Q_R) (\phi^\dagger \phi)
  + \coef{Y_2} (\bar{Q}_L \, \mathbf{T}^a_Q Q_R) (\phi^\dagger \sigma^a \phi),
\end{align}
where $\lambda^A$ are the Gell-Mann matrices, $\sigma^a$ are the Pauli matrices,
$\mathbf{T}^A_Q$ ($\mathbf{t}^a_Q$) are the generators of $SU(2)$ ($SU(3)$) in
the representation of $Q$, and the matrices $C^a$ are defined by
\begin{align}
  C^{3/2}
  &=
  \frac{1}{\sqrt{2}}
  \left(
    \begin{array}{rr}
      1 & 0 \\
      -i & 0 \\
      0 & 0
    \end{array}
  \right),
  &
  C^{1/2}
  &=
  \frac{1}{\sqrt{6}}
  \left(
    \begin{array}{rr}
      0 & 1 \\
      0 & -i \\
      -2 & 0
    \end{array}
  \right),
  \nonumber \\
  C^{-1/2}
  &=
  -\frac{1}{\sqrt{6}}
  \left(
    \begin{array}{rr}
      1 & 0 \\
      i & 0 \\
      0 & 2
    \end{array}
  \right),
  &
  C^{-3/2}
  &=
  -\frac{1}{\sqrt{2}}\left(
    \begin{array}{rr}
      0 & 1 \\
      0 & i \\
      0 & 0
    \end{array}
  \right).
  \nonumber
\end{align}
These matrices connect the quadruplet representation of $SU(2)$ with the doublet and triplet representations. The index $i$ indicates the SM fermion family and $W_W = Y_2 =
0$ for singlets. We have used the following notation for coefficients of
operators that are linear in $Q$:
$\lambda_i$ is the coefficient of $\bar{Q}q_i\phi$, $y_i$ is for
$\bar{Q}q_i \phi\phi$, and $w_i$ is for $\bar{Q}\sigma^{\mu\nu}q_i
F_{\mu\nu}$. When there is more than one possibility, the corresponding coupling
constants are differentiated by an additional subindex, which indicates the SM
field that unambiguously determines the operator. Observe that we include all
the gauge-invariant operators of dimension equal to or smaller than 5 that can
be constructed with the field content of the theory. The condition of linear
couplings is used to select the representations of the vector-like quarks, but
not to restrict their interactions in the effective theory. Note also that the
$\lambda_i$ parameters are dimensionless, whereas $y_i$, $w_i$, $Y$ and
$W$ have dimensions of inverse energy and are expected to be of order
$\Lambda^{-1}$. We will consider in this paper only couplings to the third
family of SM quarks. This choice is made to reduce the dimensionality of the
parameter space and to automatically satisfy the most stringent flavour
limits. It is also motivated by theoretical ideas in different models. This
means that $\lambda_i$, $y_i$ and $w_i$ are taken to be vanishing for
$i=1,2$. Accordingly, we simplify the name of the non-vanishing couplings in the
following way:
\begin{align}
& \lambda= \lambda_3; ~ \lambda_t=\lambda_{u3}; ~ \lambda_b=\lambda_{d3}; \nonumber \\
&y=y_3; ~~ y_t=y_{u3}; ~~y_b=y_{d3} ;\nonumber \\ 
&w=w_3; ~~ w_B=w_{B3}; ~~ w_W=w_{W3}; ~~ w_G=w_{G3}. 
\end{align}

\begin{table}
  \centering
  \begin{tabular}{ccccccc}
    \toprule
    Name & Irrep & $\bar{Q} \phi q$ & $\bar{Q} \phi \phi q$ &
    $\bar{Q} \sigma^{\mu\nu} q F_{\mu\nu}$ \\
    \midrule
    $U$ & $1_{2/3}$ & \Y & \Y & \Y \\
    $D$ & $1_{-1/3}$ & \Y & \Y & \Y \\
    $Q_1$ & $2_{1/6}$ & \Y & \Y & \Y \\
    $Q_5$ & $2_{-5/6}$ & \Y & \Y & \N \\
    $Q_7$ & $2_{7/6}$ & \Y & \Y & \N \\
    $T_1$ & $3_{-1/3}$ & \Y & \Y & \Y \\
    $T_2$ & $3_{2/3}$ & \Y & \Y & \Y \\
    $T_4$ & $3_{-4/3}$ & \N & \Y & \N \\
    $T_5$ & $3_{5/3}$ & \N & \Y & \N \\
    $F_1$ & $4_{1/6}$ & \N & \Y & \Y \\
    $F_5$ & $4_{-5/6}$ & \N & \Y & \N \\
    $F_7$ & $4_{7/6}$ & \N & \Y & \N \\
    \bottomrule
  \end{tabular}
  
  \caption{Irreps ${(2T + 1)}_Y$ under ${SU(2)}_L \times {U(1)}_Y$ and linear
    interactions of new quarks with dimension-5 linear couplings. The subscript
    in the name of each multiplet is the absolute value of the numerator of its
    hypercharge, when written as an irreducible fraction. An explicit formula
    for this integer number is
    $|2 + 4 \widetilde{T} + 3 (Y - 2/3) / (1 - \widetilde{T})|$ where
    $\widetilde{T} = T \mod{1}$.}%
  \label{tab:dim5-reps}
\end{table}

\begin{figure}
  \centering
  \includegraphics[width=0.3\textwidth]{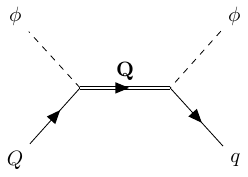}
  \qquad \qquad \qquad
  \includegraphics[width=0.3\textwidth]{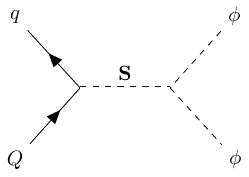}
  \caption{Tree-level diagrams that generate the $\bar{Q} q \phi \phi$ operator in UV completions of $\Lcal$ with additional extra quarks (left) and additional scalars (right).}
  \label{fig:yukawa-UV-diagrams}
\end{figure}

\begin{figure}
  \centering
  \includegraphics[width=0.5 \textwidth]{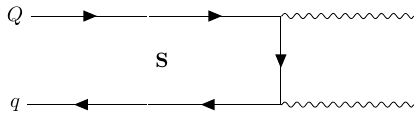}
  \caption{A one-loop diagram that generates the $\bar{Q} \sigma_{\mu\nu} q F^{\mu\nu}$
    operator in a UV completion of $\Lcal$ with new scalars.} 
  \label{fig:magnetic-UV-diagrams}
\end{figure}

Let us briefly comment on possible ultraviolet completions that can give rise to
the dimension 5 operators at low energies.  The Yukawa-like operators
$\bar{Q}q\phi\phi$, of dimension 5, can be generated at the tree level in a
completion with one additional field: either a colour-neutral scalar
$\mathbf{S}$, with interactions $\mathbf{S}\bar{Q}q$ and $\mathbf{S}\phi\phi$ or
an additional quark $\mathbf{Q}$, with interactions $\mathbf{Q}\phi Q$ and
$\mathbf{Q} \phi q$. The mass of the extra particle, which is assumed to be
larger than $M$, sets the cutoff scale $\Lambda$ of the effective theory
$\Lcal$. The Feynman diagrams that contribute to the dimension-5 Yukawas are
shown in figure~\ref{fig:yukawa-UV-diagrams}.  The quantum numbers of the extra
field must allow for the gauge-invariant vertices in the diagrams. This means
that the heavy scalar $\mathbf{S}$ belongs to one of the representations $1_0$,
$3_0$ and $3_1$ of $SU(2) \times U(1)$, while the heavy quark $\mathbf{Q}$
belongs to one of the representations in the first seven rows of
table~\ref{tab:dim5-reps}, so it is also a RVLQ (but assumed to be heavier than
the ones in the effective Lagrangian).  The operators of the form
$\bar{Q} \sigma_{\mu\nu} q F^{\mu\nu}$, on the other hand, cannot be generated
at tree-level in a renormalizable ultraviolet theory. In
figure~\ref{fig:magnetic-UV-diagrams} we show a one-loop diagram that
contributes to these effective operators in a theory with an extra scalar multiple $\mathbf{S}$, which must be either a singlet or a
  triplet of $SU(2)$, and a singlet or an octet of $SU(3)$. That is, there are 4
  possibilities: $(1, 1)_0$, $(1, 3)_0$, $(8, 1)_0$ and $(8, 3)_0$. The
coefficients $w$ of these ``magnetic'' operators are thus
naturally suppressed by a loop factor in weakly coupled completions. In
  addition, because a quark mass insertion $m_Q$ is needed for the chiralities
  of the external lines to match those of the effective operator, the
  suppression with the UV scale $m_S$ is not $1/m_S$ as expected from the
  effective theory power counting, but $m_Q/m_S^2$. An explicit model with a $U$ vector-like quark and a scalar singlet has been studied in~\cite{Kim:2018mks}. 

\section{Mixing}
\label{sec:mixing}

The multiplets in table~\ref{tab:dim5-reps} can be decomposed into component fields with
well-defined electric charge:
\begin{gather}
  Q_1 = \multiplet{T^0 \\ B^0}, \qquad
  Q_5 = \multiplet{B^0 \\ Y}, \qquad
  Q_7 = \multiplet{X \\ T^0},
  \\
  T_1 = \multiplet{T^0 \\ B^0 \\ Y}, \qquad
  T_2 = \multiplet{X \\ T^0 \\ B^0}, \qquad
  T_4 = \multiplet{B^0 \\ Y \\ Y '}, \qquad
  T_5 = \multiplet{X' \\ X \\ T^0},
  \\
  F_1 = \multiplet{X \\ T^0 \\ B^0 \\ Y}, \qquad
  F_5 = \multiplet{T^0 \\ B^0 \\ Y \\ Y'}, \qquad
  F_7 = \multiplet{X' \\ X \\ T^0 \\ B^0}.
\end{gather}
The components are denoted by symbols in the set $\{X', X, T^0, B^0, Y, Y'\}$, with electric charges given by
\begin{align}
  Q(X')  &= 8/3, & Q(B^0) &= -1/3, \\
  Q(X)   &= 5/3, & Q(Y)   &= -4/3, \\
  Q(T^0) &= 2/3, & Q(Y')  &= -7/3.
\end{align}

Upon electroweak breaking, the fields $T^0$ ($B^0$) will mix, in general, with all the Standard Model
up-type (down-type) quarks. However, with our flavour restriction and neglecting the tiny off-diagonal CKM elements of the third family, the new quarks mix only with the top and bottom quarks. The relevant mass terms have the form
\begin{align}
  \Lmass
  &=
  - \left(%
    \begin{array}{cc}
      \bar{t}^0_{L} & \bar{T}^0_L
    \end{array}
  \right)
  \left(%
    \begin{array}{cc}
      m^t_{11} & m^t_{12} \\
      m^t_{21} & m^t_{22}
    \end{array}
  \right)
  \left(%
    \begin{array}{c}
      t^0_R \\
      T^0_R
    \end{array}
  \right)
  \\
  &\phantom{=}
  -\left(%
    \begin{array}{cc}
      \bar{b}^0_L & \bar{B}^0_L
    \end{array}
  \right)
  \left(%
    \begin{array}{cc}
      m^b_{11} & m^b_{12} \\
      m^b_{21} & m^b_{22}
    \end{array}
  \right)
  \left(%
    \begin{array}{c}
      b^0_R \\
      B^0_R
    \end{array}
  \right)
  + \text{h.c.},
\end{align}
with the superindex 0 emphasizing that the fields are weak eigenstates, i.e. the
components of the gauge-covariant multiplets.\footnote{Note that we use the
  symbol $t_{R}^0$ for the right-handed SM weak eigenstate of electric charge
  3/2 (-1/3), which is in fact the unique component of the SM iso-singlet
  $u_{R3}$ ($d_{R3}$)of hypercharge 3/2 (-1/3). Of course, $t_L^0$ ($b_L^0$) are
  the upper and lower components of the SM iso-doublet $q_{L3}$.}  The elements
of the diagonal of each of the mass matrices are $m_{11} \sim v$, which arises
from the Standard Model Yukawa coupling $\bar{q} \phi q$, and $m_{22} \simeq M$. For RVLQ, one
of the off-diagonal elements, $m_{ij} \sim v$, comes from the operator
$\bar{Q} \phi q$, and the other one, $m_{ji} \sim y v^2$, comes from
$\bar{Q} \phi \phi q$. For NRVLQ, only one of the off-diagonal elements, $m_{ij} \sim y v^2$, is
non-zero. The precise values of the entries of the mass matrices are given in
table~\ref{tab:masses}. The mixing angles that
relate weak and mass eigenstates are obtained by diagonalizing the corresponding
mass matrices:
\begin{align}
  \left(
    \begin{array}{c}
      t_{L,R} \\
      T_{L,R}
    \end{array}
  \right)
  &=
  \left(%
    \begin{array}{cc}
      c^t_{L,R} & -e^{i \phi_t} s^t_{L,R} \\
      e^{-i \phi_t} s^u_{L,R} & c^t_{L,R}
    \end{array}
  \right)
  \left(%
    \begin{array}{c}
      t^0_{L,R} \\
      T^0_{L,R}
    \end{array}
  \right),
  \label{eq:mixing-u}
  \\
  \left(%
    \begin{array}{c}
      b_{L,R} \\
      B_{L,R}
    \end{array}
  \right)
  &=
  \left(%
    \begin{array}{cc}
      c^b_{L,R} & -e^{i \phi_b} s^b_{L,R} \\
      e^{-i \phi_b} s^b_{L,R} & c^b_{L,R}
    \end{array}
  \right)
  \left(%
    \begin{array}{c}
      b^0_{L,R} \\
      B^0_{L,R}
    \end{array}
  \right),
  \label{eq:mixing-d}
\end{align}
where $t$, $T$, $b$ and $B$ are the mass eigenstates,
$c^{t,b}_{L,R} := \cos\theta^{t,b}_{L,R}$ and
$s^{t,b}_{L,R} := \sin\theta^{t,b}_{L,R}$, with $\theta^{t,b}_{L,R}$ the mixing
angle.  In what follows, we take $\phi_t = \phi_b = 0$, since
non-trivial phases $\phi_{t,b}$ can be ignored for the observables discussed here.  The explicit expressions for the mixing angles in terms of $m^{t,b}_{ij}$ are
(see also ref.~\cite{Dawson:2012di})
\begin{align}
  \tan 2\theta^{t,b}_L
  &=
  \frac{
    2 \left|
      m^{t,b}_{11} {(m^{t,b}_{21})}^* + m^{t,b}_{12} {(m^{t,b}_{22})}^*
    \right|
  }{
    |m^{t,b}_{11}|^2 - |m^{t,b}_{12}|^2 - |m^{t,b}_{21}|^2 + |m^{t,b}_{22}|^2
  },
  \\
  \tan 2\theta^{t,b}_R
  &=
  \frac{
    2 \left|
      {(m^{t,b}_{11})}^* m^{t,b}_{12} + {(m^{t,b}_{21})}^* m^{t,b}_{22}
    \right|
  }{
    |m^{t,b}_{11}|^2 - |m^{t,b}_{12}|^2 - |m^{t,b}_{21}|^2 + |m^{t,b}_{22}|^2
  }.
\end{align}
From these formulas and the scale dependence of each entry it can then be seen
that, for $M\gg v$ (in agreement with experimental limits, see below), the
mixing angles are suppressed by $v/M$, at least. Furthermore,
$\theta_{L} \gg \theta_{R}$ if $|m_{12}| \gg |m_{21}|$, and viceversa. For
natural values of the couplings and $\Lambda > 1~\mathrm{TeV}$, one of the
off-diagonal couplings is indeed much larger than the other, so the off-diagonal
couplings involving heavy and light quark eigenstates will be mostly chiral
(especially in the $b$ sector).  For RVLQ, the dominant mixing angle is
$\theta_L$ for even isospin and $\theta_R$ for those with odd isospin. For
NRVLQ, instead, the dominant mixing angle is $\theta_R$ for even isospin and
$\theta_L$ for odd isospin.  Note, however, that for some RVLQ the limits from
electroweak precision tests are quite strict \cite{Aguilar-Saavedra:2013qpa}. For these
multiplets, the off-diagonal entries might be comparable and then the
interactions involving both chiralities would be relevant. 

\begin{table}
  \centering
  \begin{tabular}{ccccccc}
    \toprule
    & $m^t_{12}$ & $m^t_{21}$ & $m^t_{22}$ & $m^b_{12}$ & $m^b_{21}$ & $m^b_{22}$ \\
    \midrule
    $U$ &
    $\frac{\lambda^* v}{\sqrt{2}}$ & $\frac{y v^2}{2}$ & $\hat{M}$ &
    -- & -- & -- \\
    $D$ &
    -- & -- & -- &
    $\frac{\lambda^* v}{\sqrt{2}}$ & $\frac{y v^2}{2}$ & $\hat{M}$ \\
    \midrule
    $Q_1$ &
    $\frac{(y_u)^* v^2}{2}$ & $\frac{\lambda_u v}{\sqrt{2}}$ &
    $\hat{M} - \frac{Y_2 v^2}{4}$ &
    $\frac{(y_d)^* v^2}{2}$ & $\frac{\lambda_d v}{\sqrt{2}}$ &
    $\hat{M} + \frac{Y_2 v^2}{4}$ \\
    $Q_5$ &
    -- & -- & -- &
    $\frac{y^* v^2}{2}$ & $\frac{\lambda v}{\sqrt{2}}$ &
    $\hat{M} - \frac{Y_2 v^2}{4}$ \\
    $Q_7$ &
    $\frac{y^* v^2}{2}$ & $\frac{\lambda v}{\sqrt{2}}$ &
    $\hat{M} + \frac{Y_2 v^2}{4}$ &
    -- & -- & -- \\
    \midrule
    $T_1$ &
    $\lambda^* v$ & $\frac{y_u v^2}{\sqrt{2}}$ &
    $\hat{M} - \frac{Y_2 v^2}{2}$ &
    $-\frac{\lambda^* v}{\sqrt{2}}$ & $-\frac{y_d v^2}{2}$ &
    $\hat{M}$ \\
    $T_2$ &
    $\frac{\lambda^* v}{\sqrt{2}}$ & $-\frac{y_u v^2}{2}$ &
    $\hat{M}$ &
    $\lambda^* v$ & $\frac{y_d v^2}{\sqrt{2}}$ &
    $\hat{M} + \frac{Y_2 v^2}{2}$ \\
    $T_4$ &
    -- & -- & -- &
    0 & $\frac{y v^2}{\sqrt{2}}$ &
    $\hat{M} - \frac{Y_2 v^2}{2}$ \\
    $T_5$ &
    0 & $\frac{y  v^2}{\sqrt{2}}$ &
    $\hat{M} + \frac{Y_2 v^2}{2}$ &
    -- & -- & -- \\
    \midrule
    $F_1$ &
    $-\frac{y^* v^2}{\sqrt{6}}$ & 0 &
    $\hat{M} - \frac{Y_2 v^2}{4}$ &
    $-\frac{y^* v^2}{\sqrt{6}}$ & 0 &
    $\hat{M} + \frac{Y_2 v^2}{4}$ \\
    $F_5$ &
    $\frac{y^* v^2}{\sqrt{2}}$ & 0 &
    $\hat{M} - \frac{3 Y_2 v^2}{4}$ &
    $\frac{y^* v^2}{\sqrt{6}}$ & 0 &
    $\hat{M} - \frac{Y_2 v^2}{4}$ \\
    $F_7$ &
    $-\frac{y^* v^2}{\sqrt{6}}$ & 0 &
    $\hat{M} + \frac{Y_2 v^2}{4}$ &
    $-\frac{y^* v^2}{\sqrt{2}}$ & 0 &
    $\hat{M} + \frac{3 Y_2 v^2}{4}$\\
    \bottomrule
  \end{tabular}
  \caption{Mass matrix elements. We use the notation
    $\hat{M} = M + Y_1 v^2 /2$. The $11$ component is always just the
    Standard Model contribution:
    $m^{t,b}_{11} = \lambda^{t,b}_\SM v /\sqrt{2}$.}
  \label{tab:masses}
\end{table}

\section{Indirect effects}
\label{sec:indirect}

In this section, we discuss the indirect effects of heavy quarks in low-energy physics, Higgs physics and top physics, which are summarized in table~\ref{tab:indirect-effects}. NRVLQ typically generate smaller contributions than RVLQ, as any insertion of a dimension-5 operator introduces a suppression of $1/\Lambda$. For the same reason, the effects of the dimension-5 interactions of RVLQ will naturally be small corrections to the ones coming only from dimension-4 interactions, when they are present.

Integrating out the RVLQ at tree level gives contributions
to dimension-6 operators in the SMEFT. The low-energy effective Lagrangian, which can be read from ref.~\cite{deBlas:2017xtg}, is presented in
table~\ref{tab:SMEFT-matching}, with the corresponding effective operators
defined in table~\ref{tab:operators}. Observe that the dimension-6 terms without extra quarks in the effective theory $\Lcal$, which we are not writing here, will give additional contributions to the corresponding dimension-6 operators in the SMEFT. However, these contributions will be suppressed by  $M^2/\Lambda^2$ or $M/\Lambda$ relative to the ones from integrating out the RVLQ. Still, they might be relevant for $M/\Lambda$ not small, depending on the values of the couplings.\footnote{This is nothing but a more precise formulation of the usual caution one should exert in general with indirect bounds. Remember that we are considering the case in which only the quarks in one vector-like multiplet are lighter than $\Lambda$. In the presence of other light extra particles (including vector-like quarks) the indirect bounds would need to be reevaluated.}  Here we assume that even in this case they do not cancel against the ones in table~\ref{tab:SMEFT-matching}. 

\begin{table}[h]
  \centering
  \begin{tabular}{cccc}
    \toprule
    & Observable & Coupling & Loop order \\
    \midrule
    \multirow{3}{*}{EWPO}
    & $S$ and $T$ parameters & $\lambda_{(t)}$, $y_{(t)}$ & one loop
    \\
    & \multirow{2}{*}{$Z \to bb$} & $\lambda_{(t)}$, $y_{(t)}$ & one loop
    \\
    & & $\lambda_{(b)}$, $y_{(b)}$ & tree level
    \\
    \midrule
    \multirow{4}{*}{Higgs}
    & $H \to bb$ & $\lambda_{(b)}$, $\lambda_{(b)} y_{(b)}$ & tree level
    \\
    & $ttH$ production & $\lambda$ & tree level
    \\
    & $gg \to H$, $H \to gg$ & $\lambda_{(t)}$, $Y_1$ & one loop \\
    & double Higgs production & $\lambda_{(t)}$, $Y_1$ & one loop
    \\
    \midrule
    \multirow{3}{*}{Top}
    & top single production & $\lambda_{(t)} w_W$ & tree level
    \\
    & top pair production & $\lambda_{(t)}$, $\lambda_{(t)} w_B$ & tree level
    \\
    & $tt\gamma$ and $ttZ$ production &
    $\lambda_{(t)}$, $\lambda_{(t)} w_B$, $\lambda_{(t)} w_W$ &
    tree level
    \\
    \midrule
    low-energy $\cancel{\text{CP}}$
    & electron/neutron EDM &
    $\lambda_{(t)}$, $\lambda_{(t)} \lambda_{(b)}$, $\lambda_{(t)} w_F$
    &
    two loops
    \\
    \bottomrule
  \end{tabular}
  \caption{Summary of indirect effects of heavy quarks. The subindex $(q)$ means
    that only the couplings to the Standard Model quark $q$ should be taken. The
    dependence on products of couplings may involve complex conjugation of some
    of them.}
  \label{tab:indirect-effects}
\end{table}

\begin{table}[h]
  \centering
  \begin{tabular}{ccc}
    \toprule
    & $\Lcal_{\text{nh}}$ & $\Lcal_{\text{h}}$ \\
    \midrule
    $U$
    &
    $\frac{\lambda {(w_B)}^*}{M} \Ocal_{tB}
    + \frac{\lambda {(w_G)}^*}{M} \Ocal_{tG}
    + \left(
      \frac{{(\lambda^t_\SM)}^* {|\lambda|}^2}{2 M^2}
      + \frac{\lambda^* y}{M}
    \right) \Ocal_{t \phi}$
    &
    $\frac{{|\lambda|}^2}{4 M^2} \Ocal^{(1)}_{\phi q}
    - \frac{{|\lambda|}^2}{4 M^2} \Ocal^{(3)}_{\phi q}$
    \\
    $D$
    &
    $\frac{\lambda {(w_B)}^*}{M} \Ocal_{bB}
    + \frac{\lambda {(w_G)}^*}{M} \Ocal_{bG}
    + \left(
      \frac{{(\lambda^b_\SM)}^* {|\lambda|}^2}{2 M^2}
      + \frac{\lambda^* y}{M}
    \right) \Ocal_{b \phi}$
    &
    $- \frac{{|\lambda|}^2}{4 M^2} \Ocal^{(1)}_{\phi q}
    - \frac{{|\lambda|}^2}{4 M^2} \Ocal^{(3)}_{\phi q}$
    \\
    \midrule
    \multirow{4}{*}{$Q_1$}
    &
    $\frac{\lambda_t {(w_B)}^*}{M} \Ocal_{tB}
    + \frac{\lambda_t {(w_W)}^*}{M} \Ocal_{tW}
    + \frac{\lambda_t {(w_G)}^*}{M} \Ocal_{tG}$
    &
    \multirow{4}{*}{$\begin{array}{c}
      - \frac{{|\lambda_t|}^2}{2 M^2} \Ocal_{\phi t}
      + \frac{{|\lambda_b|}^2}{2 M^2} \Ocal_{\phi b}
      \\
      + \frac{\lambda_b {(\lambda_t)}^*}{M^2} \Ocal_{\phi tb}
    \end{array}$}
    \\
    &
    $+ \frac{\lambda_b {(w_B)}^*}{M} \Ocal_{bB}
    + \frac{\lambda_b {(w_W)}^*}{M} \Ocal_{bW}
    + \frac{\lambda_b {(w_G)}^*}{M} \Ocal_{bG}$
    &
    \\
    &
    $+\left(
      \frac{{(\lambda^t_\SM)}^* {|\lambda_t|}^2}{2 M^2}
      + \frac{\lambda_t {(y_t)}^*}{M}
      \right) \Ocal_{t \phi}$ &
    \\
    &
      $+ \left(
      \frac{{(y^b_\SM)}^* {|\lambda_b|}^2}{2 M^2}
      + \frac{\lambda_b {(y_b)}^*}{M}
    \right) \Ocal_{b \phi}$
    &
    \\
    \midrule
    $Q_5$
    &
    $\left(
      \frac{{(\lambda^b_\SM)}^* {|\lambda|}^2}{2 M^2}
      + \frac{\lambda y^*}{M}
    \right) \Ocal_{b \phi}$
    &
    $-\frac{{|\lambda|}^2}{2 M^2} \Ocal_{\phi b}$
    \\
    $Q_7$
    &
    $\left(
      \frac{{(\lambda^t_\SM)}^* {|\lambda|}^2}{2 M^2}
      + \frac{\lambda y^*}{M}
    \right) \Ocal_{t \phi}$
    &
    $\frac{{|\lambda|}^2}{2 M^2} \Ocal_{\phi t}$
    \\
    \midrule
    \multirow{2}{*}{$T_1$}
    &
    $\frac{\lambda {(w_W)}^*}{M} \Ocal_{bW}
    + \left(
      \frac{{(y^t_\SM)}^* {|\lambda|}^2}{4 M^2}
      + \frac{\lambda^* y_t}{M}
    \right) \Ocal_{t \phi}$
    &
    \multirow{2}{*}{$- \frac{3 {|\lambda|}^2}{16 M^2} \Ocal^{(1)}_{\phi q}
      + \frac{{|\lambda|}^2}{16 M^2} \Ocal^{(3)}_{\phi q}$}
    \\
    &
    $+ \left(
      \frac{{(y^b_\SM)}^* {|\lambda|}^2}{8 M^2}
      + \frac{\lambda^* y_b}{2 M}
    \right) \Ocal_{b \phi}$
    &
    \\
    \multirow{2}{*}{$T_2$}
    &
    $\frac{\lambda {(w_W)}^*}{M} \Ocal_{tW}
    + \left(
      \frac{{(y^t_\SM)}^* {|\lambda|}^2}{8 M^2}
      - \frac{\lambda^* y_b}{2 M}
    \right) \Ocal_{t \phi}$
    &
    \multirow{2}{*}{$\frac{3 {|\lambda|}^2}{16 M^2} \Ocal^{(1)}_{\phi q}
      + \frac{{|\lambda|}^2}{16 M^2} \Ocal^{(3)}_{\phi q}$}
    \\
    &
    $+ \left(
      \frac{{(y^b_\SM)}^* {|\lambda|}^2}{4 M^2}
      + \frac{\lambda^* y_b}{M}
    \right) \Ocal_{b \phi}$
    &
    \\
    \bottomrule
  \end{tabular}
  \caption{Dimension-6 effective Lagrangian generated by tree-level matching of
    the effective theory with each multiplet to the SMEFT. The contributions to
    Hermitian and non-Hermitian operators are separated in $\Lcal_{\text{h}}$
    and $\Lcal_{\text{nh}}$. The complete effective Lagrangian is
    $\Lcal_{\text{h}} + (\Lcal_{\text{nh}} + \hc)$. The definitions of the
    operators $\Ocal_i$ are given in table~\ref{tab:operators}.}
  \label{tab:SMEFT-matching}
\end{table}

\begin{table}[h]
  \centering
  \begin{tabular}{ccccc}
    \toprule
    Name & Operator & & Name & Operator \\
    \midrule
    $\Obphi$ & $(\phi^\dagger \phi) (\bar{q}_{L3} \phi d_{R3})$
    & &
    $\Otphi$ & $(\phi^\dagger \phi) (\bar{q}_{L3} \phi u_{R3})$
    \\
    $\Ophib$ & $(\phi^\dagger i\DLR_\mu \phi) (\bar{d}_{R3} \gamma^\mu d_{R3})$
    & &
    $\Ophit$ & $(\phi^\dagger i\DLR_\mu \phi) (\bar{u}_{R3} \gamma^\mu u_{R3})$
    \\
    $\Ophiq{1}$
    & $(\phi^\dagger i\DLR_\mu \phi) (\bar{q}_{L3} \gamma^\mu q_{L3})$
    & &
    $\Ophiq{3}$
    &
    $(\phi^\dagger i\DLR{}^{\, a}_\mu \phi) (\bar{q}_{L3} \gamma^\mu \sigma_a q_{L3})$
    \\
    $\Ophitb$
    &
    $(\tilde{\phi}^\dagger iD_\mu \phi) (\bar{u}_{R3} \gamma^\mu d_{R3})$
    \\
    $\ObB$ & $(\bar{q}_{L3} \sigma^{\mu\nu} d_{R3}) \phi \, B_{\mu\nu}$
    & &
    $\OtB$ & $(\bar{q}_{L3} \sigma^{\mu\nu} u_{R3}) \tilde{\phi} \, B_{\mu\nu}$
    \\
    $\ObW$ & $(\bar{q}_{L3} \sigma^{\mu\nu} d_{R3}) \sigma^a \phi \, W^a_{\mu\nu}$
    & &
    $\OtW$ & $(\bar{q}_{L3} \sigma^{\mu\nu} u_{R3}) \sigma^a \tilde{\phi} \, W^a_{\mu\nu}$
    \\
    $\ObG$ &
    $\frac{1}{2}
    (\bar{q}_{L3} \sigma^{\mu\nu} \lambda^A d_{R3})
    \phi \, G^A_{\mu\nu}$
    & &
    $\OtG$ &
    $\frac{1}{2}
    (\bar{q}_{L3} \sigma^{\mu\nu} \lambda^A u_{R3})
    \tilde{\phi} \, G^A_{\mu\nu}$
    \\
    \bottomrule
  \end{tabular}
  \caption{Dimension-6 operators generated by tree-level matching of
    the effective theory with each multiplet to the SMEFT. This basis of
    operators was proposed in~\cite{AguilarSaavedra:2009mx} and is a subset of
    the Warsaw basis~\cite{Grzadkowski:2010es}.}
  \label{tab:operators}
\end{table}

On the other hand, the NRVLQ do not contribute at tree level to the dimension-6 SMEFT. Therefore, their indirect effects are small. Their leading tree-level
contributions of NRVLQ have at least dimension 8 and will not be written explicitly.

\subsubsection*{Electroweak precision observables}

Electroweak precision observables set the strongest limits on the
Yukawa couplings of each multiplet. In the mass-eigenstate basis, the mixing between the Standard Model $b$
quark and the $B$ component of a given multiplet induces a modification of the
$Zbb$ coupling, which affects the $R_b$, $A^b_{FB}$, $A_b$ and $R_c$ observables
at tree level. $t$--$T$ mixing changes the $Ztt$ coupling. Insertions of this
modified interaction in diagrams with loops of the top quark also generate
corrections to these observables, as well as to the $S$ and $T$ parameters.

For the renormalizable multiplets, the origin of these effects can be easily identified
in the unbroken phase. They come from tree-level and one-loop diagrams
containing the $\Ocal_{\phi q}$-type operators generated by tree level
matching. Notice that the non-renormalizable multiplets will also have
contributions to these observables, but to obtain them one needs to keep
dimension-8 operators, which indicates that their effects will be smaller.

In ref.~\cite{Aguilar-Saavedra:2013qpa}, the limits on the mixing angles from electroweak
precision observables were computed, assuming renormalizability. The corrections from dimension-5 interactions can be neglected for
RVLQ. However, for NRVLQ,
the dimension-5 contribution is the leading one. Following the method in
ref.~\cite{Aguilar-Saavedra:2013qpa}, we can use the experimental measurements of $R_b$,
$A^b_{FB}$, $A_b$ and $R_c$ to obtain the following bounds: $s_L < 0.13$ for the
triplet $T_4$, $s^d_L < 0.02$ for the quadruplet $F_7$ and $s^d_L < 0.03$ for
the quadruplet $F_5$. These limits are already satisfied by the mixing angles
\begin{equation}
  \theta \sim \frac{y v^2}{M} \lesssim 0.02,
\end{equation}
for $y \leq {(\SI{3}{TeV})}^{-1}$, $M \geq \SI{1}{TeV}$.
The quadruplet $F_1$ produces a $Zbb$
coupling with an extra suppression of $m_b/M$, so it is even less constrained.
The limits from $S$ and $T$ are weaker than the ones from $Z \to bb$ when there
is a $B$ component in the multiplet. The only multiplet without such component
among the non-renormalizable ones is $T_5 \sim 3_{5/3}$. In this case both the
limits from $Z \to bb$ and from $S$ and $T$ may be relevant. 
Anyway, since these effects are loop suppressed, as long as
$y/M \leq {(\SI{1.7}{TeV})}^{-2}$,  this multiplet
satisfies these constraints.

\subsubsection*{Higgs physics}

The $\Otphi$ operator introduces a modification of the top Yukawa coupling,
which can be measured using $ttH$ production. This process has been observed at
the LHC~\cite{Aaboud:2018urx,Sirunyan:2018hoz}. The current uncertainty for the top Yukawa
coupling is however too large for the effects of $\Otphi$ to be relevant. The
situation could improve in future experiments~\cite{Cepeda:2019klc}.

The presence of $\Otphi$ also changes gluon fusion Higgs production, through its
appearance in diagrams with loops of the top quark.  In addition, there are contributions to
$gg \to H$ from the heavy-quark loops. At the renormalizable level, the contribution of the $T$ loops
is cancelled quite precisely by the effect of $t$ loops with insertions of $\Otphi$ (such cancellation does
not happen for $B$ loops)~\cite{Aguilar-Saavedra:2013qpa}. In the presence of $Qq\phi\phi$ operators the cancellation is spoiled by the contributions to $\Otphi$ proportional to $\lambda y$. However, this contribution is suppressed not only by $M/\Lambda$ but also by the small mixing. The dimension-5 interactions with $Y_1$
give yet another contribution to this process (see also ref.~\cite{Fajfer:2013wca}). This can be computed by one-loop
matching to the SMEFT. The relevant part of the effective Lagrangian is
\begin{equation}
  \mathcal{L}_{\text{1-loop}}
  \supset
  \frac{(2T + 1) \alpha_s \re(Y_1)}{12 \pi M} \OphiG,
\end{equation}
where $\Ocal_{\phi G} = \phi^\dagger \phi \, G^A_{\mu\nu} G^{A,\mu\nu}$. As we can see, the coefficient of the induced operator is not suppressed by the mixing. Bounds
on the coefficient on this operator have been calculated in
ref.~\cite{Ellis:2018gqa}. They can be translated into limits for the parameters of our theory:
\begin{equation}
  \frac{|\re(Y_1)|}{M} < \frac{1}{(2 T + 1) {(\SI{1.25}{TeV})}^2},
\end{equation}
where $T$ is the isospin of the corresponding multiplet. Of course, both $\OphiG$ and
$\Otphi$ contribute to the $H \to gg$ partial width, through tree-level and
one-loop diagrams, respectively. This is discussed in detail in
ref.~\cite{Aguilar-Saavedra:2013qpa}. These operators modify also double Higgs production, which
has not been observed yet but could be measured at the HL-LHC~\cite{Atlas:2019qfx}. Similarly, there are loop contributions to other vector-boson decay modes of the Higgs. 

On the other hand, the $H \to bb$ decay channel is modified at the tree level by the operator
$\Obphi$. Because the contribution to this operator from dimension-4 couplings
is suppressed by the Yukawa coupling of the bottom quark, while the
dimension-5 contribution does not contain this suppression, it is possible that
the dimension-5 interaction dominates. Using the limit on the coefficient of
$\Obphi$ from ref.~\cite{Ellis:2018gqa} (with milder flavour assumptions), we find the bound 
$|y_{(b)}|/M \lesssim {(\SI{0.2}{TeV})}^{-2}$.

\subsubsection*{Top physics}

Several of the dimension-6 SMEFT operators generated at tree level are relevant
for the production of the top quark. $\OtW$ and $\Ophiq{3}$ contribute to single
production, whereas $\OtG$ contributes to pair production~\cite{AguilarSaavedra:2010zi}. In
ref.~\cite{Buckley:2015lku}, upper limits on the coefficients of these operators
are derived. They range from approximately ${(\SI{0.5}{TeV})}^{-2}$ to
${(\SI{0.8}{TeV})}^{-2}$. Again, the natural values of these coefficients in our
case, which are given by $\sim \lambda^2/2 M$ and $\sim \lambda w/M$, already
satisfy these limits. The same happens for the operators $\OtB$, $\Ophit$ and
$\Ophiq{1}$, which contribute to $tt\gamma$ and $ttZ$ production, and have even
weaker limits.

\subsubsection*{Low-energy CP violation}

The imaginary part of the coefficients of the operators $\Otphi$, $\Ophitb$,
$\OtW$, $\OtB$, $\ObW$ and $\OtG$ affects the electric dipole moment of the
electron and the neutron. These low-energy observables must be computed by
performing the RG running of the coefficients down to the electroweak scale and
integrating out the top quark. In ref.~\cite{Cirigliano:2016njn,Cirigliano:2016nyn,AguilarSaavedra:2018nen}, strong
limits on the imaginary part of the coefficients have been obtained, ranging
from ${(\SI{2}{TeV})}^{-2}$ to ${(\SI{42}{TeV})}^{-2}$. Our UV parameters enter
these coefficients with the combination $\lambda w^*/M$, so either their
absolute value is very small, or all their phases must be almost equal. A
trivial way of satisfying these limits is by imposing that all parameters are
real.

\section{Production at the LHC}
\label{sec:production}

\begin{figure}[h]
  \begin{subfigure}{0.55\linewidth}
    \centering
    \includegraphics{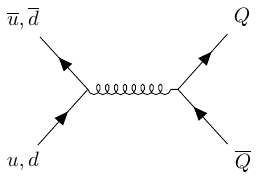}
    \caption{}\label{fig:pair-production}
  \end{subfigure}
  \begin{subfigure}{0.45\linewidth}
    \centering
    \includegraphics{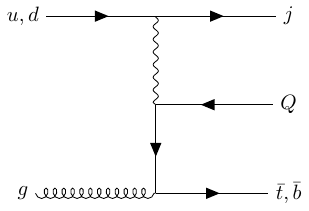}
    \caption{}\label{fig:single-production}
  \end{subfigure}
  
  \caption{Production of heavy quarks in hadron colliders: (a) example diagram
    for pair production; (b) single production in association with a light jet
    $j$ and a heavy Standard Model quark $q = t, b$.}
\end{figure}

All the vector-like quarks can be produced in pairs at hadron colliders by their coupling to gluons, which is determined by the value of $\alpha_s$ at the relevant energy. Given $M$, the production cross
section is fixed and it is the same for all the multiplets. One of the several tree-level diagrams contributing to pair production is represented in figure~\ref{fig:pair-production}. On the other hand,
the $T$, $B$ states can be singly produced via their mixing with the SM $t$, $b$ quarks. The corresponding process is represented in
figure~\ref{fig:single-production}.

\begin{figure}[h]
  \centering
  \includegraphics[width=\textwidth]{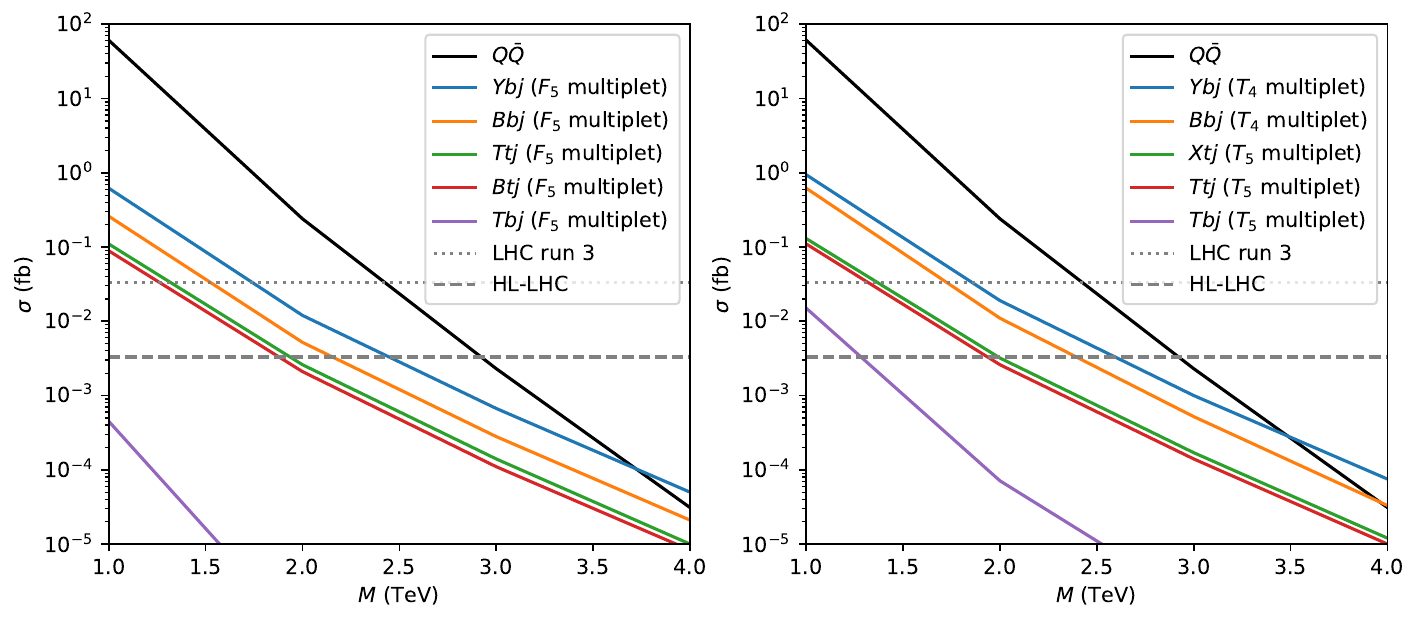}
  \caption{%
    Cross section for different processes for production of heavy
    quarks with $y = {(\SI{4}{TeV})}^{-1}$ and a center-of-mass energy
    of $\SI{14}{TeV}$. The left plot corresponds to the $F_5$
    quadruplet, while the right plot is for the $T_5$ and $T_4$
    triplets. Pair production dominates for masses below
    $\simeq\SI{3.5}{TeV}$. The dotted and dashed gray lines represent
    the minimum cross section needed to obtain at least 10 events at
    the corresponding collider, assuming that the expected integrated
    luminosity is reached~\cite{Apollinari:2116337}.  }%
  \label{fig:production-cross-sections}
\end{figure}

When the heavy quarks have low mass, the cross section for pair production is
larger than the one for single production. As their mass increases, and for fixed collider energy, the later
eventually becomes the main production mechanism. This has been studied for RVLQ mixing 
with the third family in ref.~\cite{Aguilar-Saavedra:2013qpa}. 
For these multiplets, the addition of dimension-5 interactions with natural values of the $y$ couplings and $\Lambda \geq 2$~TeV does not change significantly the results, as they give a small correction to the cross section. Here we are assuming that the dimension-4 couplings saturate the electroweak limits. In the case of NRVLQ, for natural values of the $y$ couplings and $\Lambda \geq 2$~TeV, pair production is larger than single production 
for the range of masses that can be tested at colliders in the present and near
future. Some examples of the dependence of the production cross section on the
the mass are shown in figure~\ref{fig:production-cross-sections}. 

\begin{figure}
  \centering
  \includegraphics[width=0.8\textwidth]{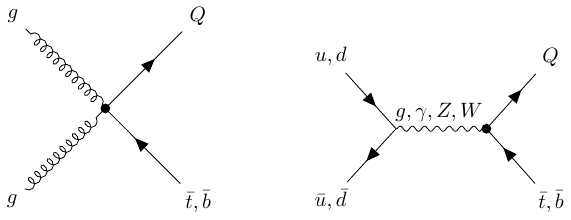}
  \caption{Single production with $\overline{Q} \sigma^{\mu\nu} q F_{\mu\nu}$-type operators.}
  \label{fig:magnetic-production}
\end{figure}

\begin{figure}
  \centering
  \includegraphics[width=0.8\textwidth]{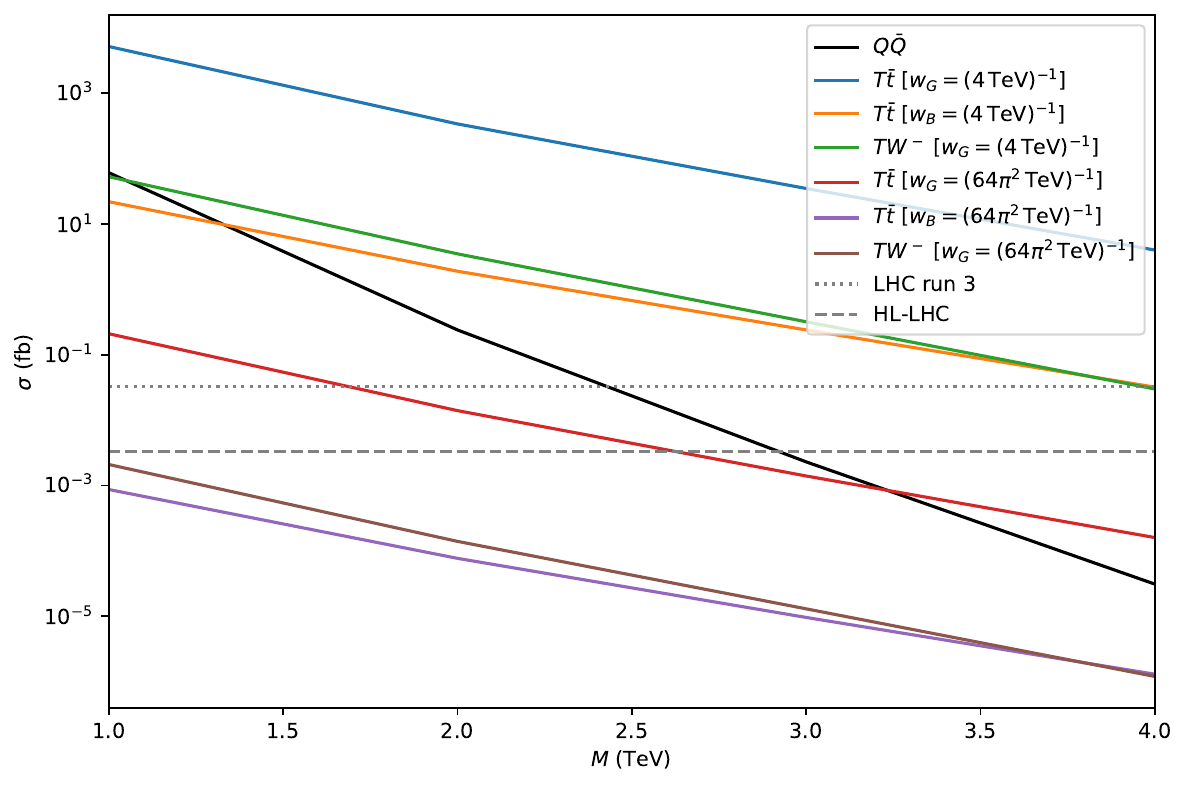}
  \caption{Cross section for different processes involving
    $\overline{Q}\sigma^{\mu\nu}q F_{\mu\nu}$, for production of heavy
    quarks in the $U$ model, with a center-of-mass energy of
    $\SI{14}{TeV}$.}
  \label{fig:production-cross-sections-magnetic}
\end{figure}

The operators $\overline{Q} \sigma^{\mu\nu} q F_{\mu\nu}$ open new single production channels, which are suppressed by $(M/\Lambda)^2$ instead of $\sin^2 \theta$. In figure~\ref{fig:magnetic-production}, we show the two main mechanisms, which produce a heavy quark in association with a Standard Model third generation quark. Other single production processes are possible with $b$ quarks from the protons in the initial state . In this way, the $B$ component of multiplets with these operators can be generated alone, while the $T$ component can be produced together with a jet or a $W$ boson. As an example, we show in figure~\ref{fig:production-cross-sections-magnetic} the cross section of the $T$ production processes involving these operators, for the $U$ multiplet. For $w = {(\SI{4}{TeV})}^{-1}$ these cross sections are large. However, these couplings are generated in renormalizable UV completions only at one loop, so the natural value for $w$ is expected to have a suppression of $1/16\pi^2$ in weakly coupled UV completions. Including this suppression gives cross sections that are smaller than pair production.

A concrete model with $\overline{Q} \sigma^{\mu\nu} q F_{\mu\nu}$
operators has been tested experimentally, as presented in
ref.~\cite{Sirunyan:2017fho}, for the case of the multiplet
$Q_1$. This analysis focuses on a particular direction in parameter
space, which in our notation corresponds to:
$g_s w^G = g w^W = -g' w^B / 6$, with the coefficients of all the
other operators set to zero. The search is for the decay into
$\gamma b$. Under these conditions, $M$-dependent limits over the
coefficients of the operators have been obtained, for masses between
$M = \SI{1}{TeV}$ and $M = \SI{1.8}{TeV}$. Translated into our
notation, the bounds for these two masses are
$w^G \lesssim {(\SI{7}{TeV})}^{-1}$ and
$w^G \lesssim {(\SI{5}{TeV})}^{-1}$, respectively.

\section{Decay}
\label{sec:decay}

In this section, we study the decays of the heavy quarks. Barring cancellations with other heavy physics, electroweak precision tests require small mixings. In this case, the splittings between the different components of the extra quark multiplet are small (of a few GeV at most for masses below 2~TeV). This in turn implies that the decays from one component to another are very suppressed. The $T$ and $B$ states can decay via mixing into $Ht$, $Zt$, $W^+b$ and $Hb$, $Zb$, $W^-b$, respectively. They can also decay into $t\gamma$, $t g$ and $b\gamma$, $b g$, respectively, in the presence of $w$ couplings. The $X$ and $Y$ states decay via mixing mainly into $W^+ t$ and $W^- b$, respectively. Their three-body decays are also sizable. Finally, $X'$ and $Y'$ have no two-body decays, as their charges differ by at least two units from the ones of the SM quarks.
 
\begin{figure}[ht]
  \centering
  \includegraphics[width=\textwidth]{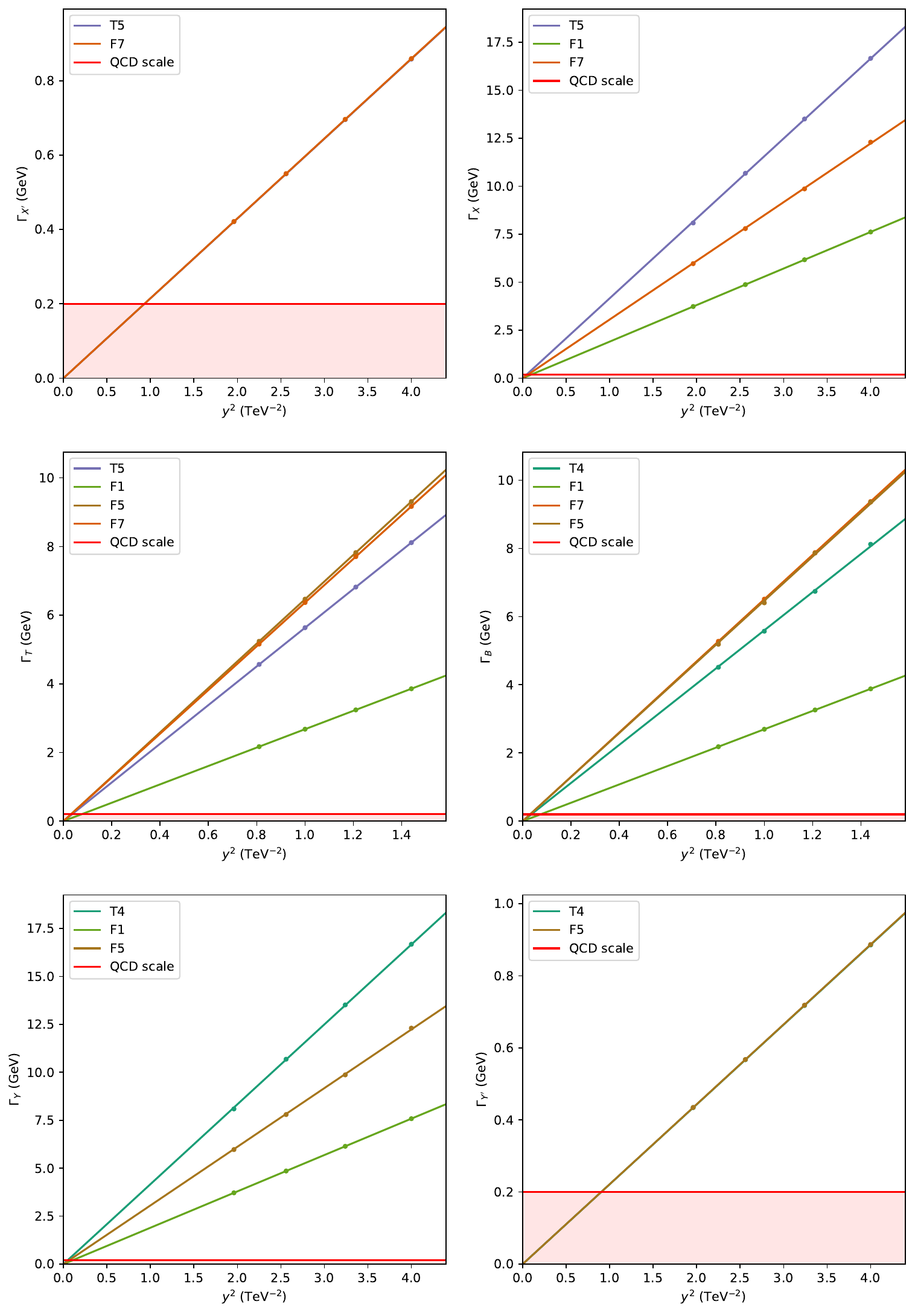}
  \caption{%
    Total decay width of $T$ (left) and $B$ (right) vs the dimension-5 Yukawa
    coupling $y$ for each multiplet without dimension-4 couplings and
    $M_Q = \SI{2}{TeV}$.%
  }
  \label{fig:widths}
\end{figure}

The decay width of RVLQ is typically large enough for them to have prompt decays and small enough for a good narrow width approximation. The NRVLQ, on the other hand, have smaller and smaller widths for larger and larger values of the cutoff $\Lambda$. In figure~\ref{fig:widths}, we show the dependence of the total width of $T$ and $B$ with the
dimension-5 Yukawa coupling $y$ for each type of NRVLQ, for $M=2~\mathrm{TeV}$. For widths below the QCD scale (see the discussion below), we have extrapolated the results calculated for larger couplings.

For small enough widths, i.e. long lifetimes, the phenomenology of the vector-like quarks can be completely different from the one in the standard searches of these particles. 
First, when the width is smaller than the QCD scale $\Lambda_{\mathrm{QCD}}$, non-perturbative effects, including hadronization, will be significant before the quarks have time to decay. One possibility is the formation near threshold of $Q\bar{Q}$ quarkonium states. This has been studied in ref.~\cite{Bigi:1986jk} (see also the review in ref.~\cite{Buchkremer:2012dn}) and generalized in ref.~\cite{Kats:2012ym} to higher color representations. Possible signatures would have di-photon and di-lepton resonant final states. But the production cross-section is suppressed by the wave function at the origin and the cross sections are small. For instance, for $M$ above  at the  0.01~TeV, ref.~\cite{Kats:2012ym} shows that the cross section into $\gamma \gamma$ for quarks with masses above 1~TeV is below 0.01~fb. In fact, most of the time the heavy quarks will fragment independently forming $Qq$ meson states and also baryons with light quarks from the vacuum. This is completely analogous to the case of $b$-quarks forming $B$ mesons. For $M \gg \Lambda_{\mathrm{QCD}}$, the mass and partial decay widths of the hadrons will inherit the properties of the heavy quark, up to small QCD corrections. Moreover, most of the energy resides in the hadron containing the heavy quark, leaving only a small fraction to light particles in the accompanying jet, and gluon radiation only softens the spectrum slightly~\cite{Bigi:1986jk}.  
Hence, the standard type of search for vectorlike quarks will be mostly blind to the fact that the quarks hadronize, as long as they decay promptly (that is, for lifetimes below $10^{-14}$~s).

For widths smaller than $\sim 10^{-12}$~GeV, the hadrons carrying the heavy quark will be long-lived. In this context, they are called $R$-hadrons. Their phenomenology at the LHC has been studied in detail, especially for squarks and gluinos in supersymmetric models. $R$-hadrons interact hadronically as they move through the detector, but in these processes the heavy quark acts mostly as a spectator of the low-energy scattering of light partons. Compared to SM hadrons, their energy loss in the calorimeter is small.
Possible signatures include (see ref.~\cite{Lee:2018pag} for a review of the phenomenology of long-lived particles):
\begin{itemize}
\item Tracks with anomalous ionization, from the slower speed of the heavy quarks in comparison to SM particles and/or non-standard charges. Note that $Qq$ mesons formed with $X$, $X'$, $Y$ or $Y'$ will always be charged, while those with $T$ and $B$ can be charged or neutral. In these searches, one must take into account the fact that the charge of the $R$-hadrons may change due to the hadronic interactions of the light partons with the detector material. 
\item Delayed detector signals, due again to the small speed. In the extreme case, it is possible for a quasi-stable $R$-hadron to loose all its energy and stop at the hadronic calorimeter; its eventual decay would give out-of-time signals.
\item Displaced vertices from the delayed decay of the heavy quark. The final states produced by $R$-hadrons with vector-like quarks are very different from the ones in supersymmetric theories and other scenarios considered thus far. So, a dedicated search for displaced vertices of vector-like quarks would be necessary to probe this scenario.
\end{itemize}
The relevance of each of the signatures depends crucially on the lifetime of the $R$-hadron, which is of the order of the lifetime of the heavy quark, as calculated ignoring QCD. In table~\ref{tab:qcd-displaced}, we give the values of $1/y$ above which i) non-perturbative QCD is important ($\LambdaQCD$), ii) displaced vertices can be observed ($\LambdaD$) and iii) the heavy quark is stable within detector distances ($\LambdaLL$).

\begin{table}[ht]
  \centering
  \begin{tabular}{cccccccccccccc}
    \toprule
    &
    \multicolumn{5}{c}{$\LambdaQCD$} &
    \multirow{2}{*}{$\LambdaD$} &
    \multirow{2}{*}{$\LambdaLL$} \\
    & $T_4$ & $T_5$ & $F_1$ & $F_5$ & $F_7$ & & \\
    \midrule
    $X'$ & -- & 1.0 & -- & -- & 1.0 & $5 \times 10^5$ & $5 \times 10^7$ \\
    $X$ & -- & 4.6 & 3.1 & -- & 3.9 & $10^6$ & $10^8$ \\
    $T$ & -- & 5.3 & 3.7 & 5.7 & 5.6 & $10^6$ & $10^8$ \\
    $B$ & 5.3 & -- & 3.7 & 5.7 & 5.7 & $10^6$ & $10^8$ \\
    $Y$ & 4.6 & -- & 3.1 & 3.9 & -- & $10^6$ & $10^8$ \\
    $Y'$ & 1.1 & -- & -- & 1.1 & -- & $5 \times 10^5$ & $5 \times 10^7$ \\
    \bottomrule
  \end{tabular}
  \caption{Value of $1/y$ (in $\si{TeV}$) at which the total width reaches the
    scales $\LambdaQCD = \SI{0.2}{GeV}$, $\LambdaD = \SI{e-12}{GeV}$ and
    $\LambdaLL = \SI{e-16}{GeV}$. For $\LambdaD$ and $\LambdaLL$ only an
    estimate of the order of magnitude is provided, obtained by
    extrapolation of the results above $\LambdaQCD$.}
  \label{tab:qcd-displaced}
\end{table}

As a reference, ATLAS has recently put bounds on the mass of long-lived supersymmetric $R$-hadrons, using ionization energy loss and time-of-flight information~\cite{Aaboud:2019trc}. This search is quite model-independent and can be adapted to the case of vector-like quarks (which are also color-triplets but fermions, rather than scalars). Comparing with the limits on production cross sections for squarks and sgluinos, we estimate a lower bound close to 1500~GeV on the mass of detector stable vector-like quarks. 

In the following we concentrate on branching ratios, having in mind mostly the case with prompt decays. 
Consider RVLQ. If the dimension-5
couplings are turned off, $T$ essentially decays only into $Ht$, $Zt$ or $W^+b$, while
$B$ decays into $Hb$, $Zt$ or $W^-t$. Changing the specific values of the
parameters in these models has a small effect in the branching ratios. This
means that the branching ratios are approximately determined by the choice
of multiplet. Because the sum of branching ratios must be one
\begin{equation}
  \label{eq:dim4-brs}
  BR(Q \to Hq) + BR(Q \to Zq) + BR(Q \to W^{\pm}q') = 1,
\end{equation}
(with $Q = T,B$ and $q,q' = t,b$) it suffices to know two branching ratios of
$Q$ to be able to know the third. Any two branching ratios $BR_1$ and $BR_2$ of
$Q$ form a point in the triangle $BR_1 + BR_2 \leq 1$, $BR_{1,2} \geq 0$. Thus,
each multiplet determines a point in this triangle (or a short segment, taking
into account variations of the values of the parameters). This is the usual
method for representing graphically the branching ratios of 
vector-like quarks~\cite{Aguilar-Saavedra:2013pxa}.

\begin{figure}[h]
  \centering
  \includegraphics[width=10cm]{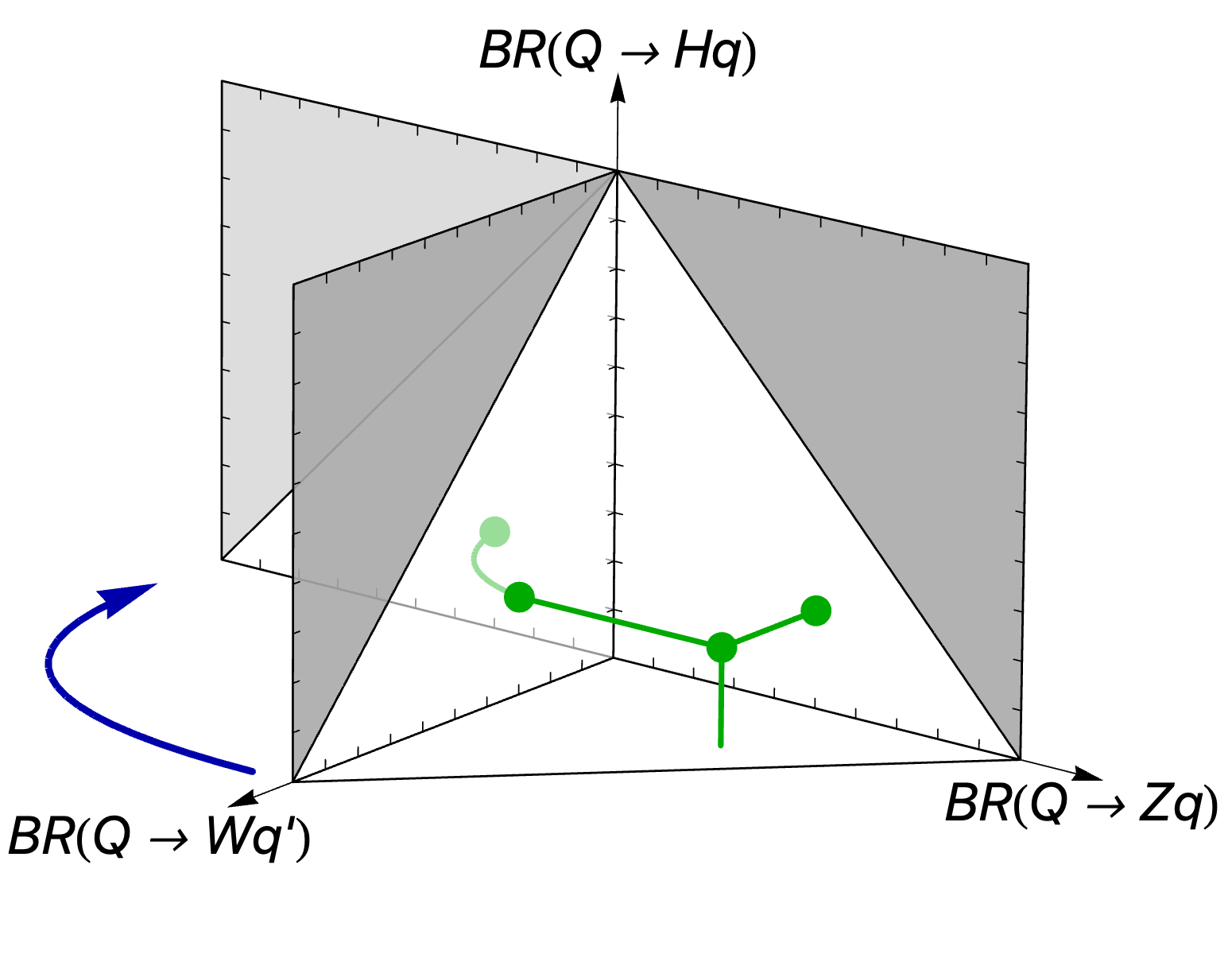}
  \caption{Representation of the
    $\left(BR(Q \to Zq), BR(Q \to W^{\pm}q'), BR(Q \to Hq)\right)$
    point as its projections into the $BR(Q \to Zq)$---$BR(Q \to Hq)$
    plane and into the $BR(Q \to W^{\pm}q')$---$BR(Q \to Hq)$ plane.}%
  \label{fig:brs-3d}
\end{figure}

The addition of dimension-5 interactions modifies these points, both by
changing the corresponding partial widths and by introducing new decay channels.
Then, eq.~\eqref{eq:dim4-brs} no longer holds. For any choice of the values of
the parameters, the branching ratios define a point $p$ in the multi-dimensional
simplex determined by $BR_i \leq 1$, $\sum_i BR_i = 1$. In particular, the
branching ratios into $Ht$, $Zt$ and $W^+b$ define a point that falls inside the
tetrahedron
\begin{align}
  \Sigma := BR(Q \to Hq) + BR(Q \to Zq) + BR(Q \to W^{\pm}q') &\leq 1, \\
  BR(Q \to Hq), BR(Q \to Zq), BR(Q \to W^{\pm}q') &\geq 0.
\end{align}
For their graphical representation, we have chosen to plot the projections of
$p$ into the $BR(Q \to Zq)$---$BR(Q \to Hq)$ plane and into the
$BR(Q \to W^{\pm}q')$---$BR(Q \to Hq)$ plane, as shown in figure~\ref{fig:brs-3d}.

\begin{figure}[h]
  \centering
  \includegraphics[width=0.9\textwidth]{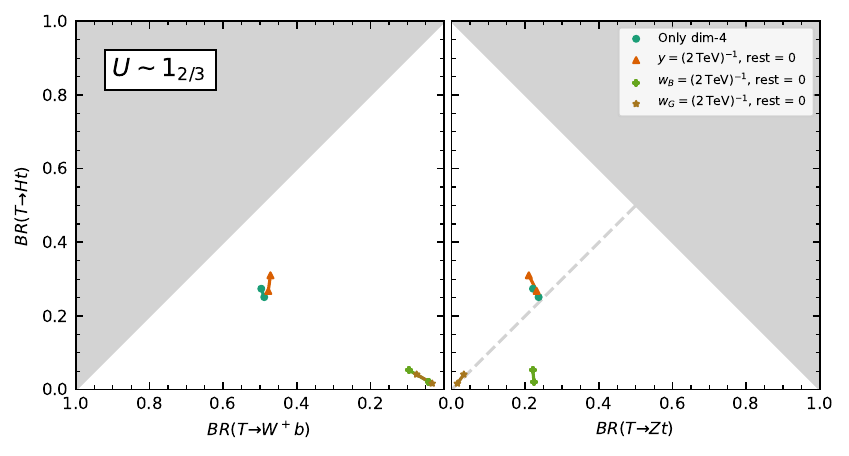}
  \includegraphics[width=0.9\textwidth]{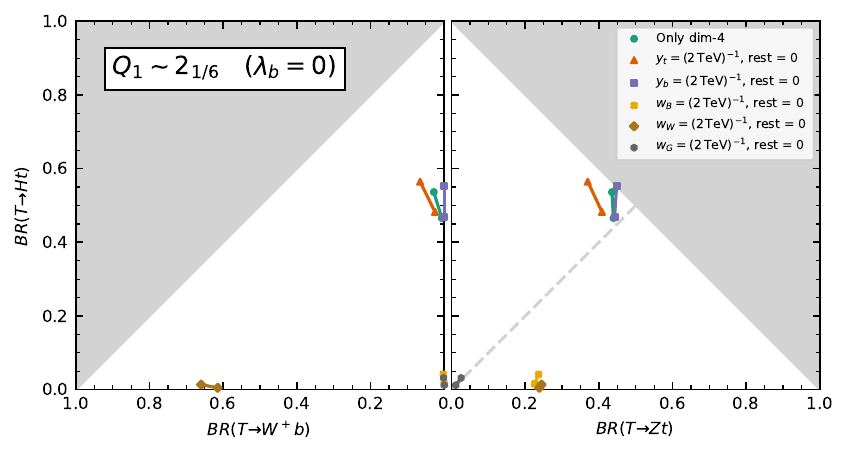}
  \includegraphics[width=0.9\textwidth]{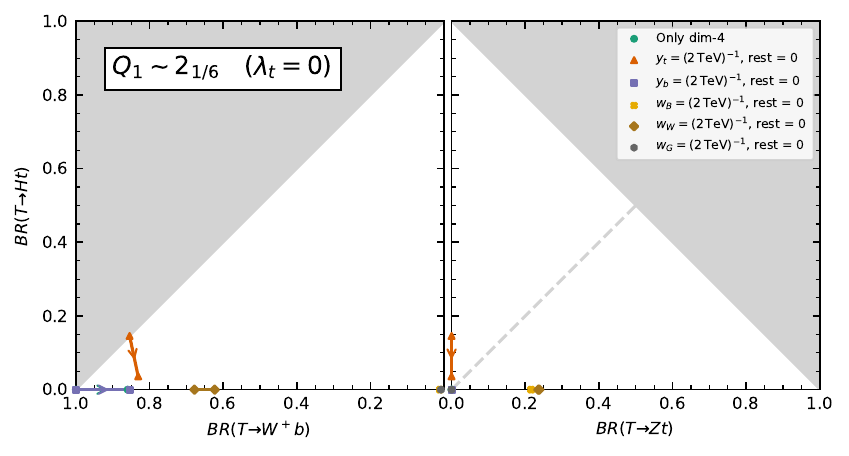}
  \caption{Branching ratios of $T$ into $Ht$, $Zt$ and $W^+b$ for various values
    of the parameters in the $U$, $Q_7$ and $Q_1$ models.  The
    dimensionless couplings $\lambda$ are always chosen to saturate the corresponding electroweak precision bounds.}%
  \label{fig:brs-u-q7-q1-th}
\end{figure}

\begin{figure}[h]
  \centering
  \includegraphics[width=0.9\textwidth]{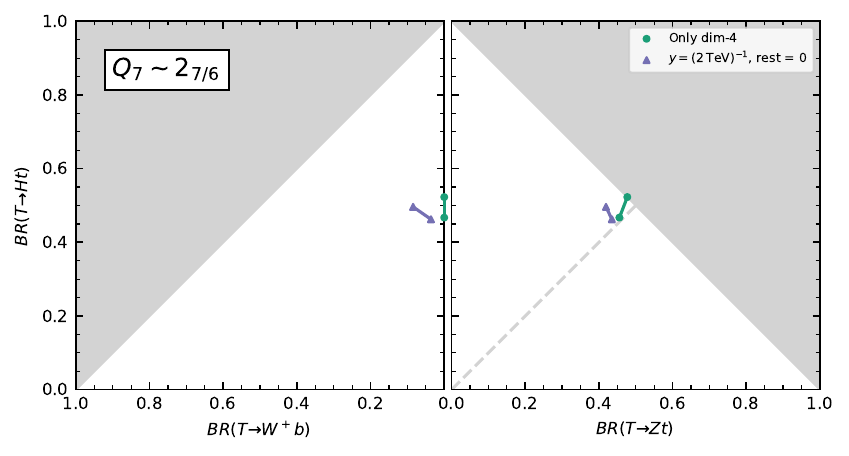}
  \includegraphics[width=0.9\textwidth]{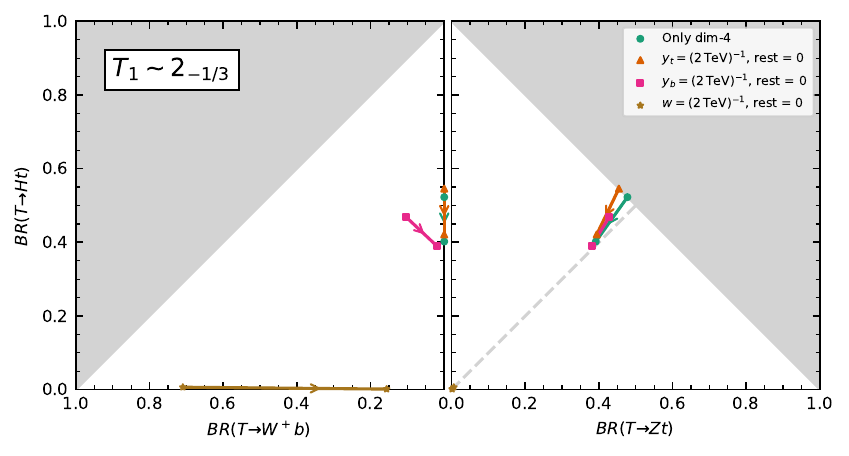}
  \includegraphics[width=0.9\textwidth]{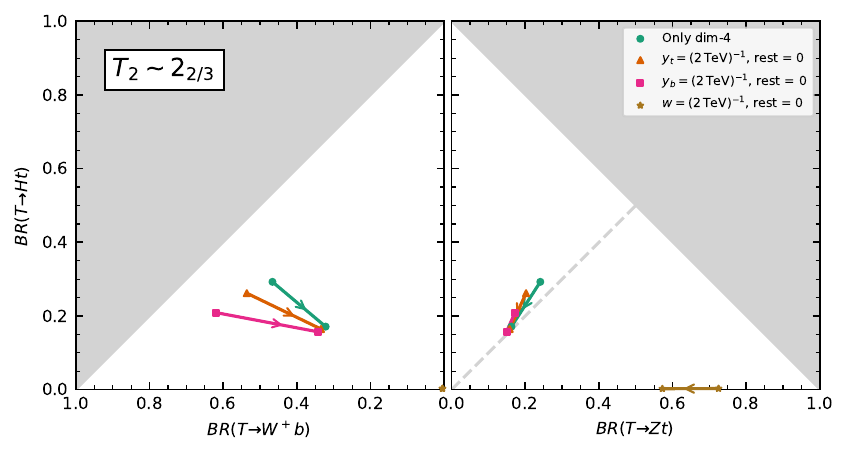}
  \caption{Branching ratios of $T$ into $Ht$, $Zt$ and $W^+b$ for various values
    of the parameters in the $Q_1$, $T_2$ and $T_1$ models.  The
    dimensionless couplings $\lambda$ are always chosen to saturate the corresponding electroweak precision bounds.}%
  \label{fig:brs-q1-t2-t1-th}
\end{figure}

\begin{figure}[h]
  \centering
  \includegraphics[width=0.9\textwidth]{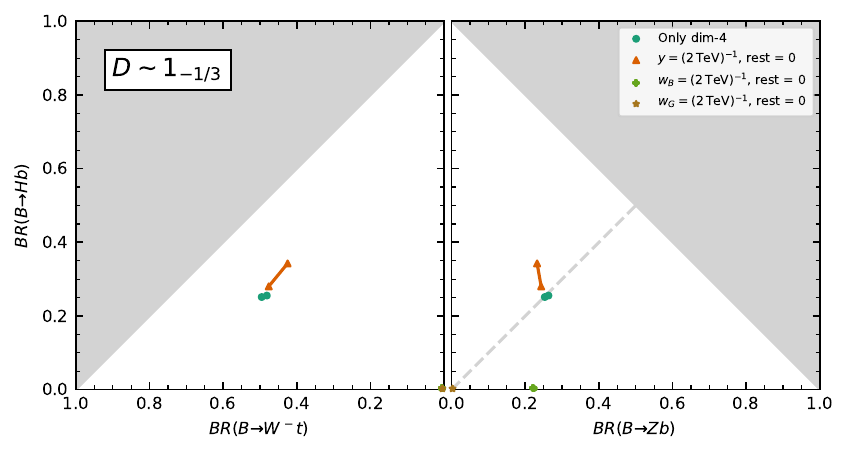}
  \includegraphics[width=0.9\textwidth]{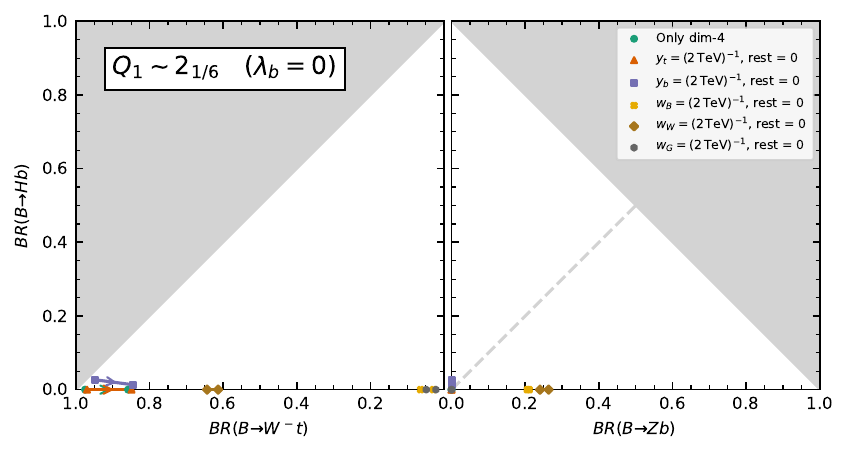}
  \includegraphics[width=0.9\textwidth]{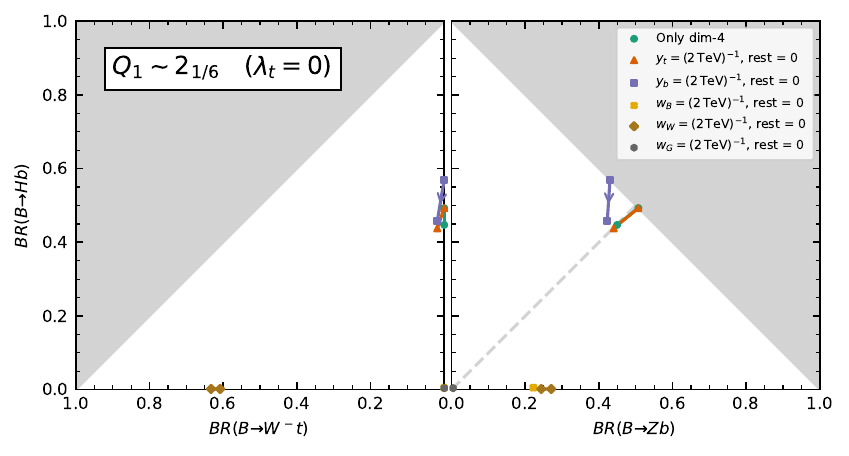}
  \caption{Branching ratios of $B$ into $Hb$, $Zb$ and $W^-t$ for various values
    of the parameters in the $D$ and $Q_1$ models.  The
    dimensionless couplings $\lambda$ are always chosen to saturate the corresponding electroweak precision bounds.}%
  \label{fig:brs-d-q1-bh}
\end{figure}

\begin{figure}[h]
  \centering
  \includegraphics[width=0.9\textwidth]{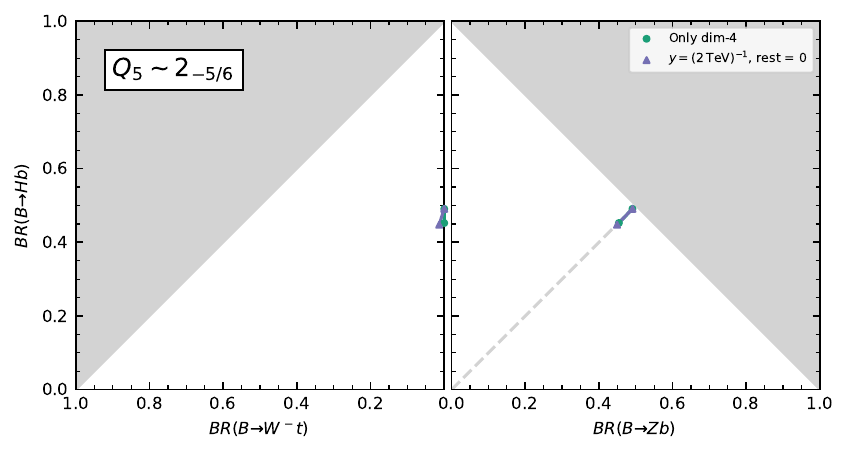}
  \includegraphics[width=0.9\textwidth]{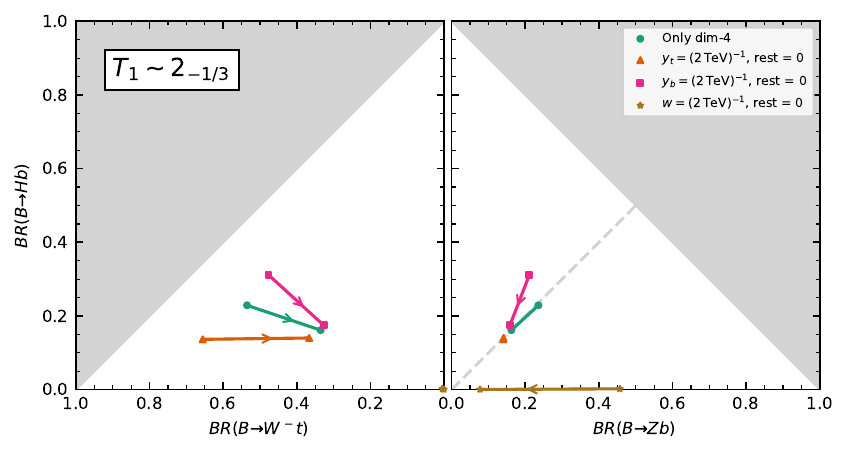}
  \includegraphics[width=0.9\textwidth]{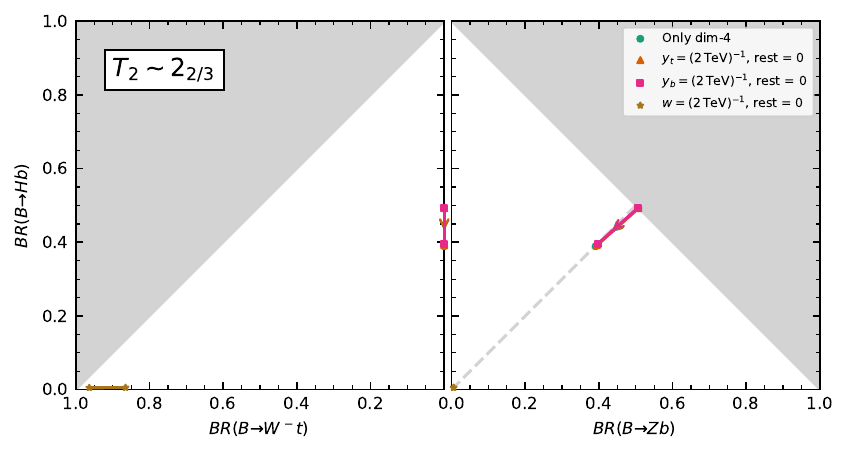}
  \caption{Branching ratios of $B$ into $Hb$, $Zb$ and $W^-t$ for various values
    of the parameters in the $Q_5$, $T_2$ and $T_1$ models.  The
    dimensionless couplings $\lambda$ are always chosen to saturate the corresponding electroweak precision bounds.}%
  \label{fig:brs-q5-t2-t1-bh}
\end{figure}

\begin{figure}[h]
  \centering
  \includegraphics[width=0.9\textwidth]{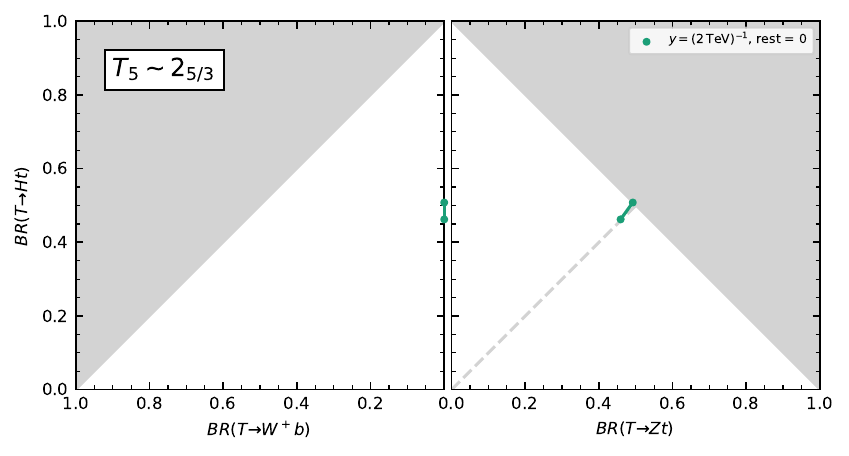}
  \includegraphics[width=0.9\textwidth]{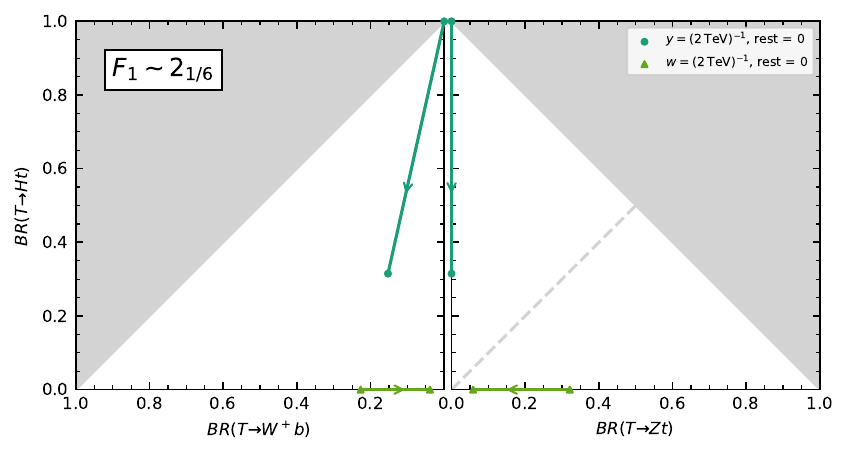}
  \caption{Branching ratios of $T$ into $Ht$, $Zt$ and $W^+b$ for various values
    of the parameters in the $T_5$ and $F_7$ models.}%
  \label{fig:brs-t5-f7-th}
\end{figure}

\begin{figure}[h]
  \centering
  \includegraphics[width=0.9\textwidth]{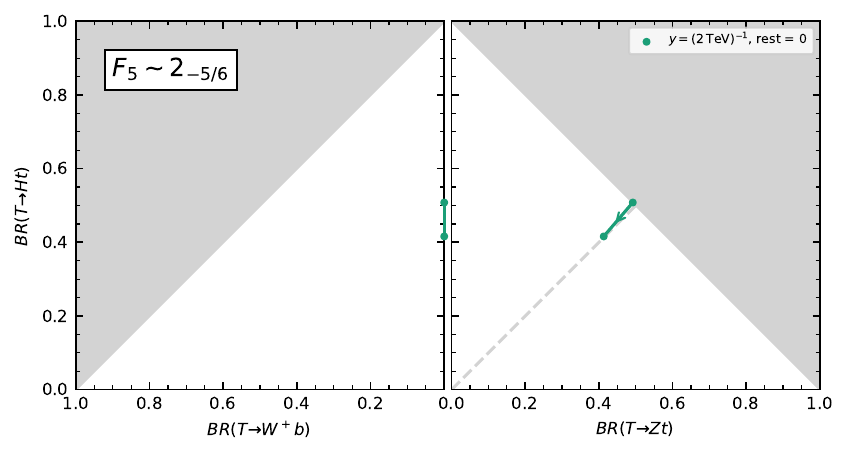}
  \includegraphics[width=0.9\textwidth]{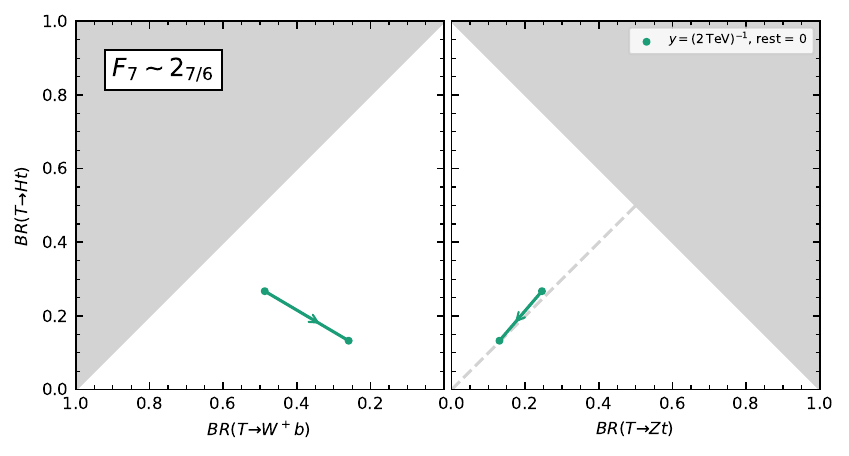}
  \caption{Branching ratios of $T$ into $Ht$, $Zt$ and $W^+b$ for various values
    of the parameters in the $F_1$ and $F_5$ models.}%
  \label{fig:brs-f1-f5-th}
\end{figure}

\begin{figure}[h]
  \centering
  \includegraphics[width=0.9\textwidth]{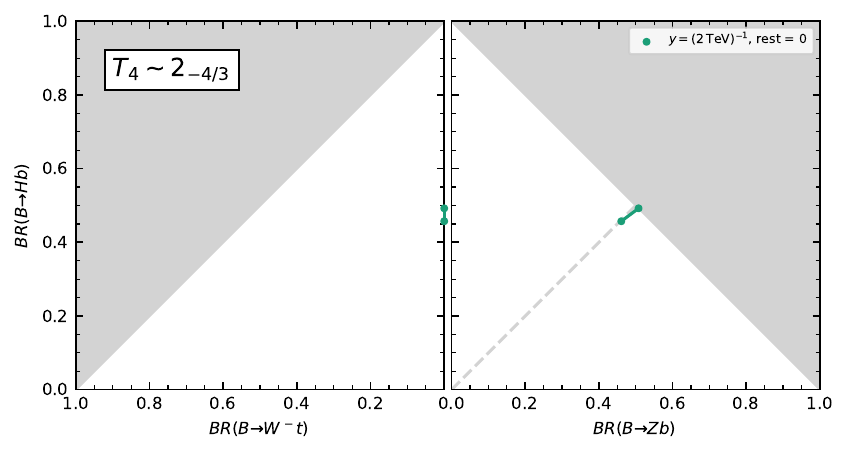}
  \includegraphics[width=0.9\textwidth]{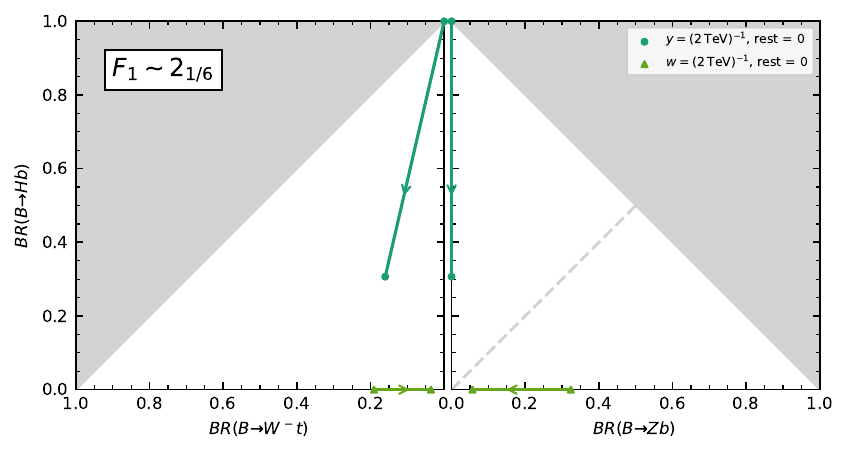}
  \caption{Branching ratios of $B$ into $Hb$, $Zb$ and $W^-t$ for various values
    of the parameters in the $T_4$ and $F_7$ models.}%
  \label{fig:brs-t4-f7-bh}
\end{figure}

\begin{figure}[h]
  \centering
  \includegraphics[width=0.9\textwidth]{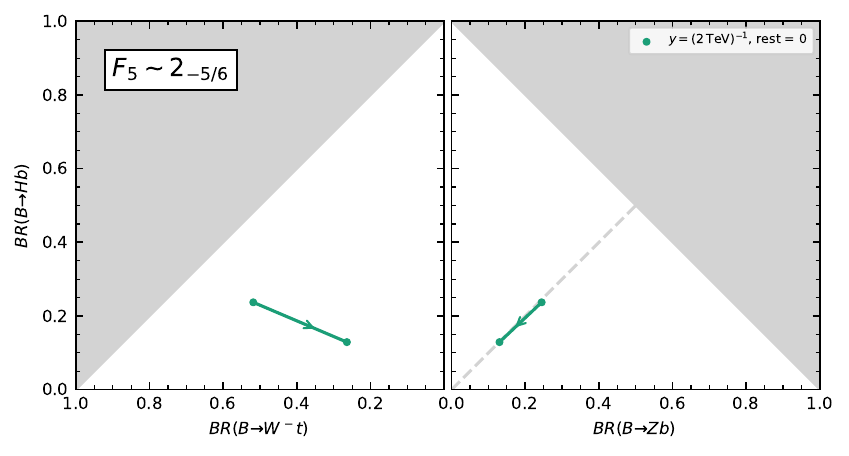}
  \includegraphics[width=0.9\textwidth]{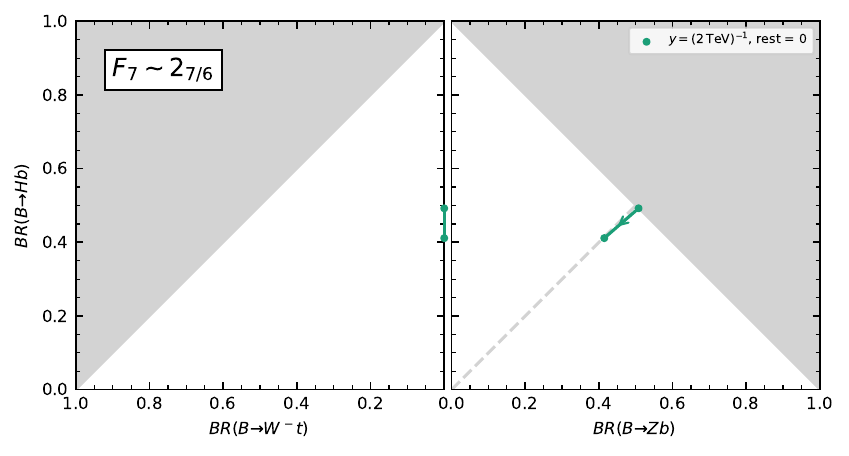}
  \caption{Branching ratios of $T$ into $Ht$, $Zt$ and $W^-t$ for various values
    of the parameters in the $F_1$ and $F_5$ models.}%
  \label{fig:brs-f1-f5-bh}
\end{figure}

The results for RVLQ are presented in figures~\ref{fig:brs-u-q7-q1-th}, \ref{fig:brs-q1-t2-t1-th}, \ref{fig:brs-d-q1-bh} and \ref{fig:brs-q5-t2-t1-bh}, while 
the branching ratios of NRVLQ are presented
in figures~\ref{fig:brs-t5-f7-th}, \ref{fig:brs-f1-f5-th}, \ref{fig:brs-t4-f7-bh} and \ref{fig:brs-f1-f5-bh}. Each
segment is obtained by evaluating at $M = \SI{1}{TeV}$ and at $M = \SI{2}{TeV}$
while keeping all the other parameters fixed. The value of the coefficients of
dimension-5 operators is chosen to be ${(\SI{2}{TeV})}^{-1}$. This pretty large value has been
chosen to visually highlight the directions of the corrections induced on the
branching ratios for RVLQ. For lower, probably
more realistic values of the coefficients (especially for $w$), these corrections will be
smaller. For the multiplets without dimension-4 interactions, this value of the
coefficients ensures that the decay width is much higher than the QCD scale, so
that that QCD effects can be neglected. The branching
ratios do not change much with the value of the corresponding coefficient in the
range from ${(\SI{2}{TeV})}^{-1}$ down to the values in which the total width equals $\LambdaQCD$. As it can be clearly seen in the figures, most branching ratios points lie near or directly over the
$BR(Q \to Hq) = BR(Q \to Zq)$ diagonal. This happens in all cases where the
coefficients of the $\bar{Q} \sigma^{\mu\nu} q F_{\mu\nu}$-type operators
vanish, except for the $F_1$ multiplet. An explanation for this fact is given in
appendix~\ref{app:diagonal}.

\begin{figure}
  \centering
  \includegraphics[width=\textwidth]{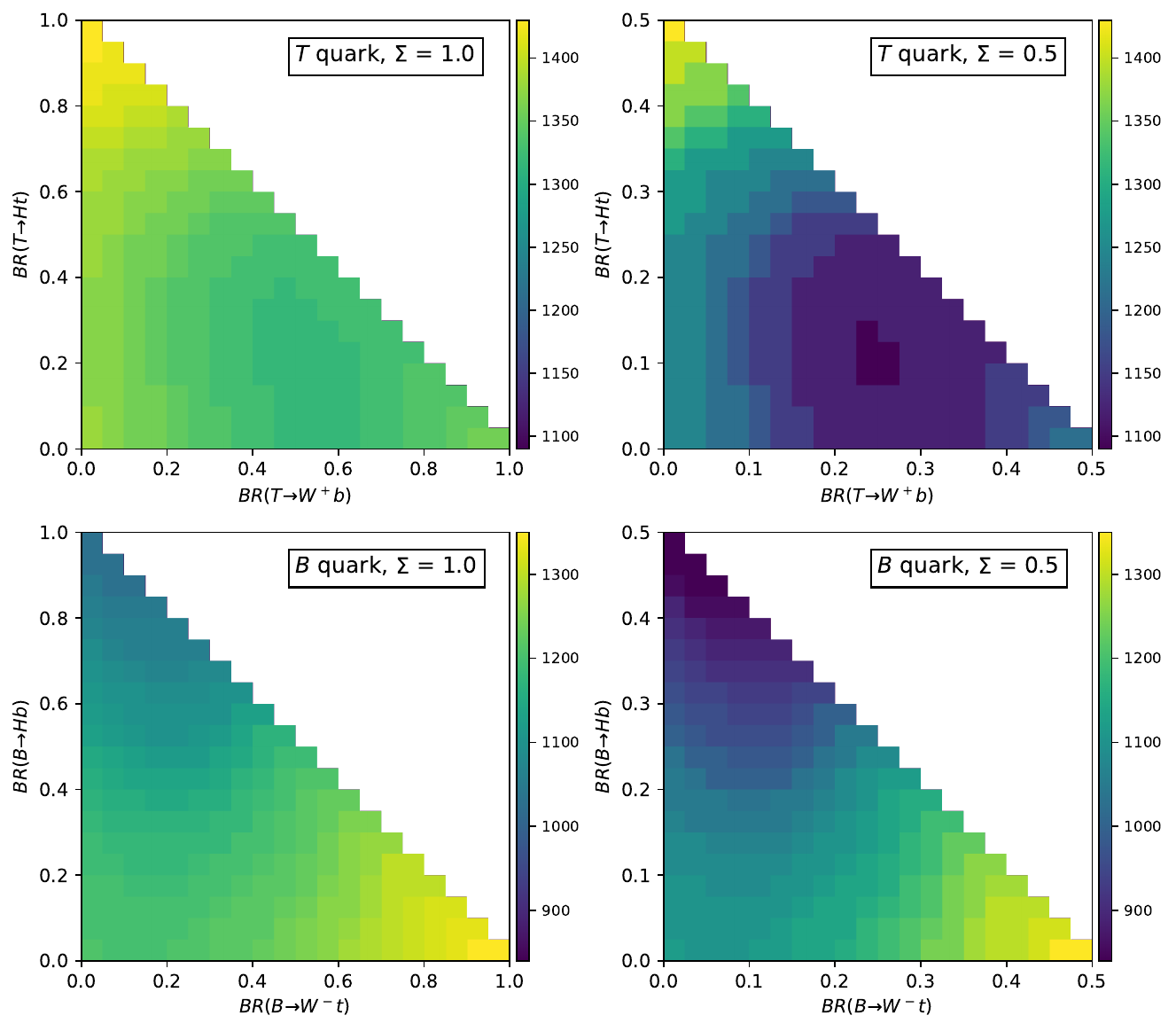}
  \caption{Left plots: lower bounds for the masses of heavy quarks presented in
    ref.~\cite{Aaboud:2018pii} assuming that the sum of branching ratios into
    $Hq$, $Zq$ and $W^\pm q'$ is $\Sigma = 1$. Right plots: corrected lower
    bounds for the case in which $\Sigma = 0.5$.}
  \label{fig:mass-limits}
\end{figure}

The experimental analyses of searches of pair-produced vector-like quarks usually combine the information on the different final states to put lower bounds on the heavy quark masses, as a function of the branching ratios to $Wq'$, $Zq$ and
$Hq$~\cite{Aaboud:2018pii,Sirunyan:2019sza}. Eq.~\eqref{eq:dim4-brs} is assumed in
these analyses, so the results are not directly valid beyond the renormalizable level. However, they can be adapted to the case where other decay
channels are present. This has been discussed previously in
refs.~\cite{Chala:2017xgc,Aguilar-Saavedra:2017giu}. In appendix~\ref{app:mass-limits} we
derive a simple formula for the corrected mass limit due to the presence of
extra decays:
\begin{equation}
  M_\Sigma = {(M_1^{1/2} + f^{1/2} \log\Sigma)}^2,
\end{equation}
It gives a lower bound $M_\Sigma$ on the mass of any heavy quark as a
function of the lower bound $M_1$ it would have if its branching
ratios into $Ht$, $Zq$ and $W^\pm q'$ were rescaled by the same factor
$1/\Sigma$, such that eq.~\eqref{eq:dim4-brs} holds. Here,
$f = \SI{20.5}{GeV}$ is just a constant. In
table~\ref{tab:mass-limits}, we present the limits calculated using
this formula, for different choices of the values of the parameters
for each model, taking the bound $M_1$ from
ref.~\cite{Aaboud:2018pii}. In all cases the couplings of dimension-4
operators are chosen to saturate the electroweak limits. In
figure~\ref{fig:mass-limits}, we show the corrections induced by the
use of this formula on the results of ref.~\cite{Aaboud:2018pii}, for
the value $\Sigma = 1/2$.

\begin{table}
  \centering
  \begin{tabular}{cc}
    \multicolumn{2}{c}{\large $U \sim 1_{2/3}$} \\
    \hline
    Only dim. 4 & 1300                     \\
    $y$        & 1310                     \\
    $w_B$      & 1010                      \\
    $w_G$      & $< 800$                  \\
    \hline
    \\
    \multicolumn{2}{c}{\large $D \sim 1_{2/3}$} \\
    \hline
    Only dim. 4 & 1200                     \\
    $y$        & 1190                     \\
    $w_B$      & $< 800$                  \\
    $w_G$      & $< 800$                  \\
    \hline
    \\
    \multicolumn{2}{c}{\large $Q_1 \sim 2_{1/6}$} \\
    \hline
    Only dim. 4  & 1340                     \\
    $y_t$      & 1340                     \\
    $y_b$      & 1120                      \\
    $w_B$      & 830                  \\
    $w_W$      & 1250                     \\
    $w_G$      & $< 800$                  \\
    \hline
    \\
    \multicolumn{2}{c}{\large $Q_5 \sim 2_{-5/6}$} \\
    \hline
    Only dim. 4 &                     1130 \\
    $y$        &                     1130 \\
    \hline
    \\
    \multicolumn{2}{c}{\large $Q_7 \sim 2_{7/6}$} \\
    \hline
    Only dim. 4 &                    1360 \\
    $y$        &                     1350 \\
    \hline
  \end{tabular}
  \qquad\qquad
  \begin{tabular}{cc}
    \multicolumn{2}{c}{\large $T_1 \sim 3_{-1/3}$} \\
    \hline
    Only dim. 4 & 1220                     \\
    $y_t$      & 1250                     \\
    $y_b$      & 1200                     \\
    $w$        & 970                  \\
    \hline
    \\
    \multicolumn{2}{c}{\large $T_2 \sim 3_{2/3}$} \\
    \hline
    Only dim. 4 & 1130                     \\
    $y_t$      & 1130                     \\
    $y_b$      & 1130                     \\
    $w$        & 1260                  \\
    \hline
    \\
    \multicolumn{2}{c}{\large $T_4 \sim 3_{-4/3}$} \\
    \hline
    $y$        &                     1130 \\
    \hline
    \\
    \multicolumn{2}{c}{\large $T_5 \sim 3_{5/3}$} \\
    \hline
    $y$        &                     1360 \\
    \hline
    \\
    \multicolumn{2}{c}{\large $F_1 \sim 4_{1/6}$} \\
    \hline
    $y$        & 1030                      \\
    $w$        & 1010                  \\
    \hline
    \\
    \multicolumn{2}{c}{\large $F_5 \sim 4_{-5/6}$} \\
    \hline
    $y$        &                      1200 \\
    \hline
    \\
    \multicolumn{2}{c}{\large $F_7 \sim 4_{7/6}$} \\
    \hline
    $y$        &                     1130 \\
    \hline
  \end{tabular}
  \caption{Mass limits for each multiplet and different values of the
    couplings. In the right column, a lower bound on the mass of the heavy quark
    (in $\si{TeV}$) is displayed, assuming that the corresponding coupling in
    the left column has a value of ${(\SI{2}{TeV})}^{-1}$ and the other dimensionful couplings vanish. The
    dimensionless couplings $\lambda$ are always chosen to saturate the corresponding electroweak precision bounds. 
  \label{tab:mass-limits}}
\end{table}

We have emphasized the presence of alternative decay channels at the non-renormalizable level. In tables~\ref{tab:extra-decays-T} and~\ref{tab:extra-decays-B}, we give the decay channels with branching ratio $> 0.01$ other than $Zq$, $W^\pm q'$ and
$Hq$ for $T$ and $B$, together with the maximum value they get and the interaction
that generates them. We choose the values $M=\SI{2}{TeV}$ and $w,y=(\SI{2}{TeV})^{-1}$, again quite extreme, in order to maximize these alternative branching ratios (including those of three-body decays). The tables also include three-body channels that survive when $w=y=0$. For RVLQ, the values of the couplings $\lambda$ are chosen to approximately saturate the electroweak limits. For smaller values of $\lambda$, the alternative channels will have larger branching ratios. Large branching ratios are found for channels involving ``magnetic'' operators. The reason is that the partial widths are suppressed in this case by $(M/\Lambda)^2$, in comparison with the $(v/M)^2$ suppression of the decay widths of standard channels. Note however that the value of $w$ we use is $\sim16 \pi^2$ times too high in weakly coupled completions. A detailed analysis of the decays of a $U$ vector-like quark into $t \gamma$ and $t g$ at the LHC has been performed in ref.~\cite{Alhazmi:2018whk}. 

 In the case of $X$ and $Y$, the decays into $W^+ t$ and $W^- b$, respectively, have branching ratios in the range 60--90\%. The remaining decays are into three particles, two of which are always $W^+ t$ or $W^- b$. The branching ratios for these channels are collected in tables~\ref{tab:extra-decays-X} and~\ref{tab:extra-decays-Y}. The states $X'$ and $Y'$ have only three-body decays. They always decay into $W^+ W^+ t$ and $W^- W^- b$, respectively.

\clearpage

\begin{table}
  \centering
  \begin{tabular}{cccc}
    \toprule
    Multiplet & Decay products & Maximum $BR$ & Coupling \\
    \midrule
    \multirow{4}{*}{$U$}
    & $b \bar{b} t$ & 0.02 & $\lambda$, $y$ \\
    & $t t \bar{t}$ & 0.01 & $\lambda$, $y$ \\
    & $\gamma t$ & 0.71 & $w_B$ \\
    & $g t$ & 0.93 & $w_G$ \\
    \midrule
    \multirow{6}{*}{$Q_1$}
    & $t W^+ W^-$ & 0.08 & $\lambda_t$, $y_b$ \\
    & $t t \bar{t}$ & 0.01 & $\lambda_t$, $y_t$, $y_b$ \\
    & $b H W^+$ & 0.11 & $y_b$ \\
    & $b Z W^+$ & 0.04 & $\lambda_b$, $y_t$, $y_b$ \\
    & $\gamma t$ & 0.77 & $w_B$ \\
    & $g t$ & 0.99 & $w_G$ \\
    \midrule
    \multirow{1}{*}{$Q_7$}
    & $t W^+ W^-$ & 0.08 & $\lambda$ \\
    \midrule
    \multirow{5}{*}{$T_1$}
    & $b H W^+$ & 0.10 & $y_b$ \\
    & $t W^+ W^-$ & 0.07 & $\lambda$, $y_t$, $y_b$ \\
    & $b Z W^+$ &  0.83 & $w$ \\
    & $t t \bar{t}$ & 0.01 & $\lambda$, $y_t$, $y_b$ \\
    & $b \gamma W^+$ & 0.01 & $w$ \\
    \midrule
    \multirow{5}{*}{$T_2$}
    & $b H W^+$ & 0.17 & $y_b$ \\
    & $t W^+ W^-$ & 0.25 & $w$ \\
    & $b Z W^+$ & 0.06 & $\lambda$, $y_t$, $y_b$ \\
    & $t b \bar{b}$ & 0.01 & $\lambda$, $y_t$, $y_b$ \\
    & $\gamma t$ & 0.21 & $w$ \\
    \midrule
    \multirow{1}{*}{$T_5$}
    & $t W^+ W^-$ & 0.08 & $y$ \\
    \midrule
    \multirow{4}{*}{$F_1$}
    & $b H W^+$ & 0.30 & $y$ \\
    & $b Z W^+$ & 0.23 & $w$ \\
    & $t W^+ W^-$ & 0.66 & $w$ \\
    & $\gamma t$ & 0.09 & $w$ \\
    \midrule
    \multirow{2}{*}{$F_5$}
    & $b H W^+$ & 0.10 & $y$ \\
    & $t W^+ W^-$ & 0.07 & $y$ \\
    \midrule
    \multirow{3}{*}{$F_7$}
    & $b H W^+$ & 0.28 & $y$ \\
    & $b Z W^+$ & 0.10 & $y$ \\
    & $t W^+ W^-$ & 0.09 & $y$ \\
    \bottomrule
  \end{tabular}
  \caption{Extra decay channels of $T$ with branching ratio larger
    than 0.01 for $M=\SI{2}{TeV}$, when the couplings $\lambda$ are fixed to the values
    that saturate electroweak precision limits. The last column
    displays the coupling constant which, when set to $(\SI{2}{TeV})^{-1}$,
    gives the maximum $BR$ in the corresponding channel. The
    appearance of $\lambda$ indicates that the channel in question is
    present already in the case with dimension-4 interactions only. }
  \label{tab:extra-decays-T}
\end{table}

\begin{table}
  \centering
  \begin{tabular}{cccc}
    \toprule
    Multiplet & Decay products & Maximum $BR$ & Coupling \\
    \midrule
    \multirow{2}{*}{$D$}
    & $\gamma b$ & 0.77 & $w_B$ \\
    & $g b$ & 0.99 & $w_G$ \\
    \midrule
    \multirow{6}{*}{$Q_1$}
    & $t H W^-$ & 0.12 & $y_t$ \\
    & $t Z W^-$ & 0.04 & $\lambda_t$, $y_t$, $y_b$ \\
    & $b W^+ W^-$ & 0.08 & $\lambda_b$, $y_t$, $y_b$ \\
    & $b t \bar{t}$ & 0.02 & $\lambda_t$, $y_t$, $y_b$ \\
    & $\gamma b$ & 0.77 & $w_B$ \\
    & $g b$ & 0.99 & $w_G$ \\
    \midrule
    \multirow{2}{*}{$Q_5$}
    & $b W^+ W^-$ & 0.08 & $\lambda$, $y$ \\
    & $b t \bar{t}$ & 0.01 & $\lambda$, $y$ \\
    \midrule
    \multirow{5}{*}{$T_1$}
    & $t H W^-$ & 0.19 & $y_t$ \\
    & $t Z W^-$ & 0.06 & $\lambda$, $y_t$, $y_b$ \\
    & $b t \bar{t}$ & 0.01 & $\lambda$, $y_t$, $y_b$ \\
    & $b W^+ W^-$ & 0.90 & $w$ \\
    & $\gamma b$ & 0.13 & $w$ \\
    \midrule
    \multirow{4}{*}{$T_2$}
    & $t H W^-$ & 0.10 & $\lambda$ \\
    & $b W^+ W^-$ & 0.07 & $\lambda$, $y_t$, $y_b$ \\
    & $t Z W^-$ & 0.12 & $w$ \\
    & $b t \bar{t}$ & 0.02 & $\lambda$, $y_t$, $y_b$ \\
    \midrule
    $T_4$  & $b W^+ W^-$ & 0.08 & $y$ \\
    \midrule
    \multirow{4}{*}{$F_1$}
    & $t H W^-$ & 0.30 & $y$ \\
    & $b W^+ W^-$ & 0.69 & $w$ \\
    & $\gamma b$ & 0.09 & $w$ \\
    & $t Z W^-$ & 0.20 & $w$ \\
    \midrule
    \multirow{3}{*}{$F_5$}
    & $t H W^-$ & 0.28 & $y$ \\
    & $t Z W^-$ & 0.10 & $y$ \\
    & $b W^+ W^-$ & 0.09 & $y$ \\
    \midrule
    \multirow{2}{*}{$F_7$}
    & $t H W^-$ & 0.10 & $y$ \\
    & $b W^+ W^-$ & 0.07 & $y$ \\
    \bottomrule
  \end{tabular}
  \caption{Extra decay channels of $B$ with branching ratio larger
    than 0.01 for $M=\SI{2}{TeV}$, when the couplings $\lambda$ are fixed to the values
    that saturate electroweak precision limits. The last column
    displays the coupling constant which, when set to $(\SI{2}{TeV})^{-1}$,
    gives the maximum $BR$ in the corresponding channel. The
    appearance of $\lambda$ indicates that the channel in question is
    present already in the case with dimension-4 interactions only.}
  \label{tab:extra-decays-B}
\end{table}

\begin{table}
  \centering
  \begin{tabular}{cccc}
    \toprule
    Multiplet & Decay products & Maximum $BR$ & Coupling \\
    \midrule
    \multirow{2}{*}{$Q_7$}
    & $t H W^+$ & 0.12 & $y$ \\
    & $t Z W^+$ & 0.04 & $\lambda$, $y$ \\
    \midrule
    \multirow{4}{*}{$T_2$}
    & $t H W^+$ & 0.10 & $\lambda$, $y_d$ \\
    & $b W^+ W^+$ & 0.04 & $\lambda$, $y_u$, $y_d$ \\
    & $t Z W^+$ & 0.12 & $w$ \\
    & $t t \bar{b}$ & 0.02 & $\lambda$, $y_u$, $y_d$ \\
    \midrule
    \multirow{2}{*}{$T_5$}
    & $t H W^+$ & 0.29 & $y$ \\
    & $t Z W^+$ & 0.11 & $y$ \\
    \midrule
    \multirow{2}{*}{$F_1$}
    & $t H W^+$ & 0.32 & $y$ \\
    & $t Z W^+$ & 0.82 & $w$ \\
    \midrule
    $F_7$
    & $t H W^+$ & 0.32 & $y$ \\
    \bottomrule
  \end{tabular}
  \caption{Decay channels of $X$ other than $W^+ t$ with branching ratio larger
    than 0.01 for $M=\SI{2}{TeV}$, when the couplings $\lambda$ are fixed to the values
    that saturate electroweak precision limits. The last column
    displays the coupling constant which, when set to $(\SI{2}{TeV})^{-1}$,
    gives the maximum $BR$ in the corresponding channel. The
    appearance of $\lambda$ indicates that the channel in question is
    present already in the case with dimension-4 interactions only.}
  \label{tab:extra-decays-X}
\end{table}

\begin{table}
  \centering
  \begin{tabular}{cccc}
    \toprule
    Multiplet & Decay products & Maximum $BR$ & Coupling \\
    \midrule
    \multirow{3}{*}{$Q_5$}
    & $b H W^-$ & 0.10 & $\lambda$, $y$ \\
    & $b Z W^-$ & 0.04 & $\lambda$, $y$ \\
    & $b b \bar{t}$ & 0.02 &  $\lambda$, $y$ \\
    \midrule
    \multirow{3}{*}{$T_1$}
    & $b H  W^-$ & 0.10 & $\lambda$, $y_u$, $y_d$ \\
    & $b Z W^-$ & 0.83 & $w$ \\
    & $t W^- W^-$ & 0.04 & $\lambda$, $y_u$, $y_d$ \\
    \midrule
    \multirow{2}{*}{$T_4$}
    & $b H W^-$ & 0.29 & $y$ \\
    & $b Z W^-$ & 0.11 & $y$ \\
    \midrule
    \multirow{2}{*}{$F_1$}
    & $b H W^-$ & 0.32 & $y$ \\
    & $b Z W^-$ & 0.82 & $w$ \\
    \midrule
    $F_5$
    & $b H W^-$ & 0.32 & $y$ \\
    \bottomrule
  \end{tabular}
  \caption{Decay channels of $Y$ other than $W^- b$ with branching ratio larger
    than 0.01  for $M=\SI{2}{TeV}$, when the couplings $\lambda$ are fixed to the values
    that saturate electroweak precision limits. The last column
    displays the coupling constant which, when set to $(\SI{2}{TeV})^{-1}$,
    gives the maximum $BR$ in the corresponding channel. The
    appearance of $\lambda$ indicates that the channel in question is
    present already in the case with dimension-4 interactions only.}
  \label{tab:extra-decays-Y}
\end{table}

\clearpage

\section{Conclusions}
\label{sec:conclusions}
The phenomenology of vector-like extra quarks near the TeV scale is to a large extent governed by gauge invariance and power counting. To start with, extra quarks can always be pair produced at hadron colliders by their gauge coupling to gluons. 
Once produced, they will decay into Standard Model particles if they have gauge-invariant linear interactions with them. 
At the renormalizable level, this is only possible for seven different gauge-covariant multiplets. These are the multiplets that can have Yukawa couplings with the Higgs doublet, which mix the extra quarks among themselves and with the SM ones. The latter mixing gives rise to decays into a SM quark and either $Z$, $W$ or Higgs bosons. In simple extensions with only one vector-like multiplet, these are the only significant decay modes. Furthermore, in the motivated case of exclusive mixing with the third generation, the branching ratios are fixed by the quantum numbers of the multiplet. 
The mixing is also responsible for indirect effects, mass splittings and single production. 

This simple picture can be modified in three ways (or combinations of them). First, one can consider general couplings to all the three SM generations~\cite{Branco:1986my,delAguila:2000rc,AguilarSaavedra:2002kr,Atre:2011ae,Cacciapaglia:2011fx}. This typically requires flavour symmetries to evade the strong flavour constraints. Sizable mixing with the valence quarks in the proton would increase the importance of single production~\cite{Atre:2011ae}. Second, it is possible to consider several vector-like quark multiplets, or other additional particles, like scalars or vector bosons. This may give rise to new production mechanisms~\cite{Barcelo:2011wu} or new decay modes~\cite{Chala:2017xgc,Aguilar-Saavedra:2017giu,Cacciapaglia:2019zmj}, in addition to the standard ones described above. Third, one can drop the assumption of renormalizability. This is the path we have explored in this work. We have proposed a model-independent approach that uses a general local effective Lagrangian, valid up to a cutoff scale $\Lambda$ and constructed with the SM fields and the fields that represent arbitrary new vector-like quarks. This is a faithful description of any model with new vector-like quarks, as long as the new physics not explicitly included appears at scales higher than $\Lambda$. In particular, the effective theory describes well the case of additional particles when they are heavier than $\Lambda$. 
As usual, the effective Lagrangian is defined by its expansion in inverse powers of $\Lambda$. The lowest order, formed by operators of canonical dimension $\leq 4$, corresponds to the usual renormalizable theories with extra vector-like quarks. The interactions of higher dimension give contributions to observables suppressed by powers of $\mu/\Lambda$, with $\mu=E,M,v$ the characteristic scale of the process. Even if suppressed, these interactions can be very relevant for proceses that do not exist at the renormalizable level. 

In our explicit phenomenological analysis we have worked with the effective theory for extensions of the SM with only one vector-like quark multiplet and we have truncated it at the next-to-leading order, i.e.\ at canonical dimension 5. For simplicity we have also assumed couplings to the third generation only. At this order, there are twelve irreducible representations of extra quarks that can decay into SM particles (and be singly produced). Up to field redefinition ambiguities, four new types of interactions appear at dimension 5: the Yukawa-type operators $\bar{Q}q\phi\phi$ and $\bar{Q}Q \phi\phi$ and the ``magnetic'' operators $\bar{Q} \sigma^{\mu\nu} q F_{\mu\nu}$ and $\bar{Q} \sigma_{\mu\nu}Q F_{\mu\nu}$. The latter have effects that increase with the energy of the process.    

We have distinguished two types of vector-like quarks. Those in the seven representations that allow for renormalizable linear interactions, and those in the remaining five representations. For the extra quark in the first group, and for natural values of the coupling constants, the dimension-5 interactions typically give only small corrections to the standard phenomenology of vector-like quarks. One exception is the possibility of new indirect effects in Higgs physics. Moreover, in strongly-coupled UV completions avoiding the loop suppression in the ``magnetic'' couplings, there can be new single production modes with cross sections larger than the one of pair production and also new decay modes (into $qg$, for instance) with large branching ratios. Of course, all these effects depend on the cutoff and will be negligible if $\Lambda$ is much larger than the TeV scale. 

For the quarks in the five multiplets that do not have renormalizable linear interactions (two triplets and four quadruplets), the dimension-5 operators give the leading contributions. In this case, all the indirect bounds can be easily evaded without explicit tuning of couplings, for moderate values of $\Lambda$. Pair production is still possible and the decay (possibly after hadronization) will be prompt if $\Lambda$ is not too high. Some non-standard decay modes, including three-body decays, can be sizable and the measurements of decays of $T$ and $B$ into $Zq$, $Wq$ and $Hq$ could easily give rise to new points in the corresponding triangles. In this respect, we have given a simple formula to recast the combination limits given by the ATLAS and CMS collaborations, which assume the absence of other decay channels. For $\Lambda \gtrsim 10^6$~TeV, the decays of the hadrons containing the heavy quarks will be non-prompt. The usual searches will not be sensitive to vector-like quarks in this regime, but one can instead resort to the signatures associated to coloured and charged long-lived particles. Taking advantage of these signatures would require dedicated searches of vector-like quarks, specially in the case of displaced vertices formed by their decay products.

New operators involving the extra quarks appear at yet higher orders in the $1/\Lambda$ expansion. At dimension 6 one should include four fermion operators~\cite{Dobrescu:2016pda}. In particular, the interactions of the form $qqqQ$ will give rise to new single production mechanisms, which can have observable cross sections at the LHC for $\Lambda$ of a few TeV when the couplings to the first generation are allowed. Moreover, at each order new types of vector-like quarks will be able to decay into SM particles. Their lifetime will be suppressed by the corresponding power of $M/\Lambda$. Finally, in principle it is also possible that new vector-like quarks exist in gauge representations with $T+Y+1/3 \not \in \mathbb{Z}$. They would be stable or else decay into additional stable particles. However, there are very strong constraints on the abundances of stable strongly interacting (and charged) particles, in particular from searches of rare nuclei~\cite{Hemmick:1989ns,Mohapatra:1997sc,Mohapatra:1998nd,Kusakabe:2009jt,Javorsek:2001yv}. \footnote{See, nevertheless, ref.~\cite{DeLuca:2018mzn} for comments on the robustness of such bounds and a proposal of coloured dark matter.} 


\acknowledgments{
We thank Juan Antonio Aguilar Saavedra, Miki Chala and Michelangelo Mangano for useful discussions. Our work has been supported by the Spanish MINECO project
FPA2016-78220-C3-1-P (Fondos FEDER) and the Junta de Andaluc\'ia grant
FQM101. The work of J.C.C.\ has also been supported by the Spanish MECD grant
FPU14.
}

\appendix

\section{Representations of Standard Model operators}
\label{app:sm-operators}

In this section, we obtain a constraint over the representation of any operator
constructed as a product of the Standard Model fields. A
representation of $SU(3) \times SU(2) \times U(1)$ is denoted by $(\rho, \sigma, Y)$, where
$\rho$ is the representation under $SU(3)$, $\sigma$ is the representation under
$SU(2)$ and $Y$ is the hypercharge. We define
\begin{equation}
  N(\rho, \sigma, Y) = A(\rho) + B(\sigma) + Y,
\end{equation}
with $0 \leq A(\rho), B(\sigma) < 1$ satisfying the equations
\begin{equation}
  \rho(e^{2 i \pi / 3} I) = e^{2 i \pi A(\rho)} I,
  \qquad
  \sigma(-I) = e^{2 i \pi B(\sigma)} I.
\end{equation}
The values $A$ and $B$ may take are limited: $A(\rho) \in \{0, 1/3, 2/3\}$ and
$B(\sigma) \in \{0, 1/2\}$. The set representations of $SU(3)$ is split into
three classes by $A$:
\begin{align}
  0 &= A(1) = A(8) = A(10) = A(\overline{10}) = A(27) = \ldots \\
  1/3 &= A(3) = A(\overline{6}) = A(15) = A(15') = A(24) = \ldots \\
  2/3 &= A(\overline{3}) = A(6) = A(\overline{15}) =
  A(\overline{15'}) = A(\overline{24}) = \ldots
\end{align}
while $B$ splits the set of $SU(2)$ representations in two: those with integer
spin and the others. Both $A$ and $B$ are additive under the operation of taking
tensor products of representations:
\begin{align}
  A(\rho_1 \otimes \rho_2) &= A(\rho_1) + A(\rho_2) \mod{1},
  \\
  B(\sigma_1 \otimes \sigma_2) &= B(\sigma_1) + B(\sigma_2) \mod{1}.
\end{align}

We will prove now that if
$(\rho_{\mathcal{O}}, \sigma_{\mathcal{O}}, Y_{\mathcal{O}})$ is the
representation of some operator $\mathcal{O}$ constructed with the Standard
Model fields, then
$N(\rho_{\mathcal{O}}, \sigma_{\mathcal{O}}, Y_{\mathcal{O}})$ is an
integer. First, it can be directly checked that
$N(\rho_\phi, \sigma_\phi, Y_\phi)$ is an integer for any Standard Model field
$\phi$. Now, from the additivity of $A$, $B$ and $Y$ it follows that the value
of $N$ corresponding to the product $\mathcal{O} \mathcal{Q}$ of two operators
$\mathcal{O}$ and $\mathcal{Q}$ is
\begin{align*}
  N(\rho_{\mathcal{OQ}}, \sigma_{\mathcal{OQ}}, Y_{\mathcal{OQ}})
  &=
  [A(\rho_{\mathcal{O}}) + A(\rho_{\mathcal{Q}})]
  + [B(\sigma_{\mathcal{O}}) + B(\sigma_{\mathcal{Q}})]
  + [Y_{\mathcal{O}} + Y_{\mathcal{Q}}]
  \mod{1}
  \\
  &=
  [A(\rho_{\mathcal{O}}) + B(\sigma_{\mathcal{O}}) + Y_{\mathcal{O}}]
  + [A(\rho_{\mathcal{Q}}) + B(\sigma_{\mathcal{Q}}) + Y_{\mathcal{Q}}]
  \mod{1}
  \\
  &=
  N(\rho_{\mathcal{O}}, \sigma_{\mathcal{O}}, Y_{\mathcal{O}})
  + N(\rho_{\mathcal{Q}}, \sigma_{\mathcal{Q}}, Y_{\mathcal{Q}})
  \mod{1},
\end{align*}
Therefore, if $N(\rho_{\mathcal{O}}, \sigma_{\mathcal{O}}, Y_{\mathcal{O}})$ and
$N(\rho_{\mathcal{Q}}, \sigma_{\mathcal{Q}}, Y_{\mathcal{Q}})$ are integers, then
$N(\rho_{\mathcal{OQ}}, \sigma_{\mathcal{OQ}}, Y_{\mathcal{OQ}})$ must also be
integer. This completes the proof.

Particularizing for $SU(3)$ triplets, we get
\begin{equation}
  T + Y + 1/3 \in \mathbb{Z},
\end{equation}
where $T$ is the spin of the $SU(2)$ representation.

\section{Limits on the mass for the case with extra decays}
\label{app:mass-limits}

Experimental data determines an upper limit $L_{\text{exp}}$ on the sum of the
cross-sections for the production and decay of a pair of heavy quarks, weighted
by the efficiency for each decay channel (see ref.~\cite{Aguilar-Saavedra:2017giu}):
\begin{equation}
  \label{eq:exp-xs}
  \sigmaProd(M) \sum_{ij} \epsilon_{ij} BR_i BR_j < L_{\text{exp}},
\end{equation}
where $M$ is the mass of the heavy quark, $i$ and $j$ run over all the decay
channels, and $\epsilon_{ij}$ is the corresponding efficiency. A limit on the
mass can be derived from this inequality. In the usual experimental analyses, it
is assumed that the sum of the branching ratios into these three channels is
$\Sigma = 1$.

We consider now the case $\Sigma < 1$. We will obtain a lower limit on the mass
of some heavy quark with branching ratios $BR_i$. Some assumption has to be made
about the efficiency $\epsilon_{ia} = \epsilon_{ai}$ for the channels $a$ that
are not $Hq$, $Zq$ or $W^\pm q'$. We adopt here the conservative choice
$\epsilon_{ia} = 0$. Let $M_1$ be the lower limit on the mass for the branching
ratios $BR^\Sigma_i = BR_i/\Sigma$, whose sum is 1, so that $M_1$ is known from
experimental analyses. We define the mass $M_\Sigma$ by the equation
\begin{equation}
  \label{eq:MSigma-definition}
  \Sigma^2 \sigmaProd(M_\Sigma) = \sigmaProd(M_1).
\end{equation}
Then, we have the identity
\begin{equation}
  \sigmaProd(M_\Sigma) \sum_{ij} \epsilon_{ij} BR_i BR_j
  =
  \sigmaProd(M_1) \sum_{ij} \epsilon_{ij} BR^\Sigma_i BR^\Sigma_j.
\end{equation}
Because $M_1$ is the limit obtained from eq.~\eqref{eq:exp-xs} for branching
ratios $BR_i^\Sigma$, it follows from this identity that $M_\Sigma$ is the limit
for $BR_i$. We now proceed to find an analytic solution to
eq.~\eqref{eq:MSigma-definition}. The production cross-section $\sigmaProd(M)$
can be approximated, for masses around $\widetilde{M} = \SI{1.1}{TeV}$ by an
exponential:
\begin{equation}
  \label{eq:approx-prod-xs}
  \sigmaProd(M)
  \simeq
  \sigmaProd(\widetilde{M})
  \exp\left(
    -\frac{M^{1/2} - \widetilde{M}^{1/2}}{f^{1/2}/2}
  \right), 
\end{equation}
where $f = \SI{20.5}{GeV}$.  In the range
  $[0.8, 1.4]\,\si{TeV}$, the difference between the cross section produced by
  this formula and the one obtained using MadGraph increases towards the
  extremes of the interval and is at most 3\%. Plugging eq.~\eqref{eq:approx-prod-xs} in
eq.~\eqref{eq:MSigma-definition} gives
\begin{equation}
  M_\Sigma = (M_1^{1/2} + f^{1/2} \log \Sigma)^2.
\end{equation}

\section{Approximate equality of the branching ratios to $Hq$ and $Zq$}%
\label{app:diagonal}

We provide here an explanation for the fact that the branching ratios of a heavy
quark $Q = T,B$ with only Yukawa-type couplings ($Q\phi q$ and $Q \phi \phi q$)
into $Zt$ and $Ht$ are approximately equal, for $Q$ in any multiplet except
$F_1$. We define $X^{L,R}_{Qq}$ and $Y^{L,R}_{Qq}$ as the following coefficients
in the Lagrangian:
\begin{align*}
  \mathcal{L}_Z &=
  - \frac{g}{2 c_W}
  \bar{q}
  \slashed{Z}
  \left(
    \pm X^L_{qQ} P_L \pm X^R_{qQ} P_R
  \right)
  Q
  + \text{h.c.}, \\
  \mathcal{L}_H &=
  - \frac{g m_Q}{2 m_W}
  \bar{q}
  H
  \left(
    Y^L_{qQ} P_L + Y^R_{qQ} P_R
  \right)
  Q,
\end{align*}
the equality of the braching ratios follows from the equality in magnitude of
the dominant $X^{L,R}_{Qq}$ and the dominant $Y^{L,R}_{Qq}$.

The weak eigenstates $q^0$, $Q^0$ couple to the $Z$ boson as
\begin{align*}
  \mathcal{L}_Z
  = &
  -\frac{g}{2 c_W}
  \sum_{\chi = L, R}
  \left(
    \begin{array}{cc}
      \bar{q}^0_\chi & \bar{Q}^0_\chi
    \end{array}
  \right)
  \slashed{Z}
  \left(
    \begin{array}{c}
      2 T_3(q^0_\chi) - 2 Q_e(q^0_\chi) s_W^2 \qquad \qquad 0 \\
      0 \qquad \qquad 2 T_3(Q^0_\chi) - 2 Q_e(Q^0_\chi) s_W^2
    \end{array}
  \right)
  \left(
    \begin{array}{c}
      q^0_\chi \\
      Q^0_\chi
    \end{array}
  \right),
\end{align*}
where $T_3$ denotes the third component of isospin and $Q_e$ denotes electric
charge. After the unitary transformation in equations~\eqref{eq:mixing-u}
and~\eqref{eq:mixing-d}, we get
\begin{equation*}
  X^{L,R}_{qQ} =
  2 s_{L,R} \, c_{L,R}
  \left[
    T_3(q^0_{L,R}) - T_3(Q^0_{L,R})
  \right].
\end{equation*}

On the other hand, the quark gauge eigenstates $q^0$, $Q^0$ couple to the Higgs
as
\begin{align*}
  \mathcal{L}_{H}
  = &
  -\frac{1}{\sqrt{2}}
  \left(
    \begin{array}{cc}
      \bar{q}^0_L & \bar{Q}^0_L
    \end{array}
  \right)
  H
  \left(
    \begin{array}{cc}
      y_{11} & y_{12} \\
      y_{21} & 0
    \end{array}
  \right)
  \left(
    \begin{array}{c}
      q^0_R \\
      Q^0_R
    \end{array}
  \right).
\end{align*}

Generally, one of the off-diagonal elements is negligible. This happens because
the dimension-4 and dimension-5 Yukawas always contribute to different elements
of the $y_{ij}$ matrix. Either one of them is zero or, when both are present,
the dimension-5 one is smaller. This means that one of the mixing angles
$\theta_{L,R}$ dominates. For multiplets with dimension-4 couplings, the
chirality with the dominant mixing angle $\theta_D$ is $D = L$ for singlets and
triplets and $D = R$ for doublets. For multiplets without dimension-4 couplings
it is $D = R$ for triplets and $D = L$ for quadruplets.

The dominant off-diagonal element $y_D$ is related to the corresponding mixing
angle as $y_D \simeq x \sqrt{2} m_Q s_D / v$, where $x = 1$ in the cases with
dimension-4 interactions and $x = 2$ in the ones with only dimension-5
ones. This factor is necessary because of the different the relation between the
mass and Yukawa terms in both cases. The dominant $HqQ$ coupling is, then
\begin{equation*}
  Y^D_{Qq} \simeq x s_D c_D.
\end{equation*}

For $X^D_{Qq} \simeq Y^D_{Qq}$ it is necessary and sufficient that
\begin{equation}
  \left|T_3(q^0_D) - T_3(Q^0_D)\right| = x/2.
\end{equation}
It can be checked case by case that this relation is statisfied for all
multiplets except for $F_1$. In this case, we have
$\left|T_3(q^0_L) - T_3(Q^0_L)\right| = 0$.

\bibliographystyle{JHEP.bst}
\bibliography{references}{}

\providecommand{\href}[2]{#2}\begingroup\raggedright\begin{thebibliography}{10}

\bibitem{delAguila:2008pw}
F.~del Aguila, J.~de~Blas and M.~P\'erez-Victoria, \emph{{Effects of new
  leptons in Electroweak Precision Data}},
  \href{https://doi.org/10.1103/PhysRevD.78.013010}{\emph{Phys. Rev.}
  {\bfseries D78} (2008) 013010}
  [\href{https://arxiv.org/abs/0803.4008}{{\ttfamily 0803.4008}}].

\bibitem{delAguila:2008cj}
F.~del Aguila and J.~A. Aguilar-Saavedra, \emph{{Distinguishing seesaw models
  at LHC with multi-lepton signals}},
  \href{https://doi.org/10.1016/j.nuclphysb.2008.12.029}{\emph{Nucl. Phys.}
  {\bfseries B813} (2009) 22}
  [\href{https://arxiv.org/abs/0808.2468}{{\ttfamily 0808.2468}}].

\bibitem{Fritzsch:1975sr}
H.~Fritzsch, M.~Gell-Mann and P.~Minkowski, \emph{{Vector - Like Weak Currents
  and New Elementary Fermions}},
  \href{https://doi.org/10.1016/0370-2693(75)90040-4}{\emph{Phys. Lett.}
  {\bfseries 59B} (1975) 256}.

\bibitem{Kearney:2013cca}
J.~Kearney, A.~Pierce and J.~Thaler, \emph{{Exotic Top Partners and Little
  Higgs}}, \href{https://doi.org/10.1007/JHEP10(2013)230}{\emph{JHEP}
  {\bfseries 10} (2013) 230} [\href{https://arxiv.org/abs/1306.4314}{{\ttfamily
  1306.4314}}].

\bibitem{Agashe:2004rs}
K.~Agashe, R.~Contino and A.~Pomarol, \emph{{The Minimal composite Higgs
  model}}, \href{https://doi.org/10.1016/j.nuclphysb.2005.04.035}{\emph{Nucl.
  Phys.} {\bfseries B719} (2005) 165}
  [\href{https://arxiv.org/abs/hep-ph/0412089}{{\ttfamily hep-ph/0412089}}].

\bibitem{Batra:2005rh}
P.~Batra, B.~A. Dobrescu and D.~Spivak, \emph{{Anomaly-free sets of fermions}},
  \href{https://doi.org/10.1063/1.2222081}{\emph{J. Math. Phys.} {\bfseries 47}
  (2006) 082301} [\href{https://arxiv.org/abs/hep-ph/0510181}{{\ttfamily
  hep-ph/0510181}}].

\bibitem{Kaplan:1991dc}
D.~B. Kaplan, \emph{{Flavor at SSC energies: A New mechanism for dynamically
  generated fermion masses}},
  \href{https://doi.org/10.1016/S0550-3213(05)80021-5}{\emph{Nucl. Phys.}
  {\bfseries B365} (1991) 259}.

\bibitem{Gherghetta:2000qt}
T.~Gherghetta and A.~Pomarol, \emph{{Bulk fields and supersymmetry in a slice
  of AdS}}, \href{https://doi.org/10.1016/S0550-3213(00)00392-8}{\emph{Nucl.
  Phys.} {\bfseries B586} (2000) 141}
  [\href{https://arxiv.org/abs/hep-ph/0003129}{{\ttfamily hep-ph/0003129}}].

\bibitem{Choudhury:2001hs}
D.~Choudhury, T.~M.~P. Tait and C.~E.~M. Wagner, \emph{{Beautiful mirrors and
  precision electroweak data}},
  \href{https://doi.org/10.1103/PhysRevD.65.053002}{\emph{Phys. Rev.}
  {\bfseries D65} (2002) 053002}
  [\href{https://arxiv.org/abs/hep-ph/0109097}{{\ttfamily hep-ph/0109097}}].

\bibitem{ArkaniHamed:2001nc}
N.~Arkani-Hamed, A.~G. Cohen and H.~Georgi, \emph{{Electroweak symmetry
  breaking from dimensional deconstruction}},
  \href{https://doi.org/10.1016/S0370-2693(01)00741-9}{\emph{Phys. Lett.}
  {\bfseries B513} (2001) 232}
  [\href{https://arxiv.org/abs/hep-ph/0105239}{{\ttfamily hep-ph/0105239}}].

\bibitem{Contino:2006qr}
R.~Contino, L.~Da~Rold and A.~Pomarol, \emph{{Light custodians in natural
  composite Higgs models}},
  \href{https://doi.org/10.1103/PhysRevD.75.055014}{\emph{Phys. Rev.}
  {\bfseries D75} (2007) 055014}
  [\href{https://arxiv.org/abs/hep-ph/0612048}{{\ttfamily hep-ph/0612048}}].

\bibitem{delAguila:1982fs}
F.~del Aguila and M.~J. Bowick, \emph{{The Possibility of New Fermions With
  $\Delta$ I = 0 Mass}},
  \href{https://doi.org/10.1016/0550-3213(83)90316-4}{\emph{Nucl. Phys.}
  {\bfseries B224} (1983) 107}.

\bibitem{delAguila:2000rc}
F.~del Aguila, M.~P\'erez-Victoria and J.~Santiago, \emph{{Observable
  contributions of new exotic quarks to quark mixing}},
  \href{https://doi.org/10.1088/1126-6708/2000/09/011}{\emph{JHEP} {\bfseries
  09} (2000) 011} [\href{https://arxiv.org/abs/hep-ph/0007316}{{\ttfamily
  hep-ph/0007316}}].

\bibitem{delAguila:2000aa}
F.~del Aguila, M.~P\'erez-Victoria and J.~Santiago, \emph{{Effective
  description of quark mixing}},
  \href{https://doi.org/10.1016/S0370-2693(00)01071-6}{\emph{Phys. Lett.}
  {\bfseries B492} (2000) 98}
  [\href{https://arxiv.org/abs/hep-ph/0007160}{{\ttfamily hep-ph/0007160}}].

\bibitem{Lavoura:1992np}
L.~Lavoura and J.~P. Silva, \emph{{The Oblique corrections from vector - like
  singlet and doublet quarks}},
  \href{https://doi.org/10.1103/PhysRevD.47.2046}{\emph{Phys. Rev.} {\bfseries
  D47} (1993) 2046}.

\bibitem{Carena:2006bn}
M.~Carena, E.~Ponton, J.~Santiago and C.~E.~M. Wagner, \emph{{Light Kaluza
  Klein States in Randall-Sundrum Models with Custodial SU(2)}},
  \href{https://doi.org/10.1016/j.nuclphysb.2006.10.012}{\emph{Nucl. Phys.}
  {\bfseries B759} (2006) 202}
  [\href{https://arxiv.org/abs/hep-ph/0607106}{{\ttfamily hep-ph/0607106}}].

\bibitem{Anastasiou:2009rv}
C.~Anastasiou, E.~Furlan and J.~Santiago, \emph{{Realistic Composite Higgs
  Models}}, \href{https://doi.org/10.1103/PhysRevD.79.075003}{\emph{Phys. Rev.}
  {\bfseries D79} (2009) 075003}
  [\href{https://arxiv.org/abs/0901.2117}{{\ttfamily 0901.2117}}].

\bibitem{AguilarSaavedra:2009es}
J.~A. Aguilar-Saavedra, \emph{{Identifying top partners at LHC}},
  \href{https://doi.org/10.1088/1126-6708/2009/11/030}{\emph{JHEP} {\bfseries
  11} (2009) 030} [\href{https://arxiv.org/abs/0907.3155}{{\ttfamily
  0907.3155}}].

\bibitem{Aguilar-Saavedra:2013qpa}
J.~A. Aguilar-Saavedra, R.~Benbrik, S.~Heinemeyer and M.~P\'erez-Victoria,
  \emph{{Handbook of vectorlike quarks: Mixing and single production}},
  \href{https://doi.org/10.1103/PhysRevD.88.094010}{\emph{Phys. Rev.}
  {\bfseries D88} (2013) 094010}
  [\href{https://arxiv.org/abs/1306.0572}{{\ttfamily 1306.0572}}].

\bibitem{Cacciapaglia:2010vn}
G.~Cacciapaglia, A.~Deandrea, D.~Harada and Y.~Okada, \emph{{Bounds and Decays
  of New Heavy Vector-like Top Partners}},
  \href{https://doi.org/10.1007/JHEP11(2010)159}{\emph{JHEP} {\bfseries 11}
  (2010) 159} [\href{https://arxiv.org/abs/1007.2933}{{\ttfamily 1007.2933}}].

\bibitem{Beauceron:2014ila}
S.~Beauceron, G.~Cacciapaglia, A.~Deandrea and J.~D. Ruiz-Alvarez, \emph{{Fully
  hadronic decays of a singly produced vectorlike top partner at the LHC}},
  \href{https://doi.org/10.1103/PhysRevD.90.115008}{\emph{Phys. Rev.}
  {\bfseries D90} (2014) 115008}
  [\href{https://arxiv.org/abs/1401.5979}{{\ttfamily 1401.5979}}].

\bibitem{Barducci:2014ila}
D.~Barducci, A.~Belyaev, M.~Buchkremer, G.~Cacciapaglia, A.~Deandrea,
  S.~De~Curtis et~al., \emph{{Framework for Model Independent Analyses of
  Multiple Extra Quark Scenarios}},
  \href{https://doi.org/10.1007/JHEP12(2014)080}{\emph{JHEP} {\bfseries 12}
  (2014) 080} [\href{https://arxiv.org/abs/1405.0737}{{\ttfamily 1405.0737}}].

\bibitem{Moretti:2016gkr}
S.~Moretti, D.~O'Brien, L.~Panizzi and H.~Prager, \emph{{Production of extra
  quarks at the Large Hadron Collider beyond the Narrow Width Approximation}},
  \href{https://doi.org/10.1103/PhysRevD.96.075035}{\emph{Phys. Rev.}
  {\bfseries D96} (2017) 075035}
  [\href{https://arxiv.org/abs/1603.09237}{{\ttfamily 1603.09237}}].

\bibitem{Carvalho:2018jkq}
A.~Carvalho, S.~Moretti, D.~O'Brien, L.~Panizzi and H.~Prager, \emph{{Single
  production of vectorlike quarks with large width at the Large Hadron
  Collider}}, \href{https://doi.org/10.1103/PhysRevD.98.015029}{\emph{Phys.
  Rev.} {\bfseries D98} (2018) 015029}
  [\href{https://arxiv.org/abs/1805.06402}{{\ttfamily 1805.06402}}].

\bibitem{Matsedonskyi:2012ym}
O.~Matsedonskyi, G.~Panico and A.~Wulzer, \emph{{Light Top Partners for a Light
  Composite Higgs}}, \href{https://doi.org/10.1007/JHEP01(2013)164}{\emph{JHEP}
  {\bfseries 01} (2013) 164} [\href{https://arxiv.org/abs/1204.6333}{{\ttfamily
  1204.6333}}].

\bibitem{DeSimone:2012fs}
A.~De~Simone, O.~Matsedonskyi, R.~Rattazzi and A.~Wulzer, \emph{{A First Top
  Partner Hunter's Guide}},
  \href{https://doi.org/10.1007/JHEP04(2013)004}{\emph{JHEP} {\bfseries 04}
  (2013) 004} [\href{https://arxiv.org/abs/1211.5663}{{\ttfamily 1211.5663}}].

\bibitem{Matsedonskyi:2015dns}
O.~Matsedonskyi, G.~Panico and A.~Wulzer, \emph{{Top Partners Searches and
  Composite Higgs Models}},
  \href{https://doi.org/10.1007/JHEP04(2016)003}{\emph{JHEP} {\bfseries 04}
  (2016) 003} [\href{https://arxiv.org/abs/1512.04356}{{\ttfamily
  1512.04356}}].

\bibitem{Rattazzi:2003ea}
R.~Rattazzi, \emph{{Cargese lectures on extra-dimensions}},  in \emph{{Particle
  physics and cosmology: The interface. Proceedings, NATO Advanced Study
  Institute, School, Cargese, France, August 4-16, 2003}}, pp.~461--517, 2003,
  \href{https://arxiv.org/abs/hep-ph/0607055}{{\ttfamily hep-ph/0607055}},
  \href{http://weblib.cern.ch/abstract?CERN-PH-TH-2006-029-JOURNAL-REF:-PARTICLE-PHYSICS}{http://weblib.cern.ch/abstract?CERN-PH-TH-2006-029-JOURNAL-REF:-PARTICLE-PHYSICS}.

\bibitem{Fajfer:2013wca}
S.~Fajfer, A.~Greljo, J.~F. Kamenik and I.~Mustac, \emph{{Light Higgs and
  Vector-like Quarks without Prejudice}},
  \href{https://doi.org/10.1007/JHEP07(2013)155}{\emph{JHEP} {\bfseries 07}
  (2013) 155} [\href{https://arxiv.org/abs/1304.4219}{{\ttfamily 1304.4219}}].

\bibitem{Alloul:2013bka}
A.~Alloul, N.~D. Christensen, C.~Degrande, C.~Duhr and B.~Fuks,
  \emph{{FeynRules 2.0 - A complete toolbox for tree-level phenomenology}},
  \href{https://doi.org/10.1016/j.cpc.2014.04.012}{\emph{Comput. Phys. Commun.}
  {\bfseries 185} (2014) 2250}
  [\href{https://arxiv.org/abs/1310.1921}{{\ttfamily 1310.1921}}].

\bibitem{Alwall:2011uj}
J.~Alwall, M.~Herquet, F.~Maltoni, O.~Mattelaer and T.~Stelzer, \emph{{MadGraph
  5 : Going Beyond}},
  \href{https://doi.org/10.1007/JHEP06(2011)128}{\emph{JHEP} {\bfseries 06}
  (2011) 128} [\href{https://arxiv.org/abs/1106.0522}{{\ttfamily 1106.0522}}].

\bibitem{Alwall:2014bza}
J.~Alwall, C.~Duhr, B.~Fuks, O.~Mattelaer, D.~G. Ozturk and C.-H. Shen,
  \emph{{Computing decay rates for new physics theories with FeynRules and
  MadGraph 5 \_aMC@NLO}},
  \href{https://doi.org/10.1016/j.cpc.2015.08.031}{\emph{Comput. Phys. Commun.}
  {\bfseries 197} (2015) 312}
  [\href{https://arxiv.org/abs/1402.1178}{{\ttfamily 1402.1178}}].

\bibitem{Aguilar-Saavedra:2017giu}
J.~A. Aguilar-Saavedra, D.~E. L\'opez-Fogliani and C.~Mu\~noz, \emph{{Novel
  signatures for vector-like quarks}},
  \href{https://doi.org/10.1007/JHEP06(2017)095}{\emph{JHEP} {\bfseries 06}
  (2017) 095} [\href{https://arxiv.org/abs/1705.02526}{{\ttfamily
  1705.02526}}].

\bibitem{Kats:2012ym}
Y.~Kats and M.~J. Strassler, \emph{{Probing Colored Particles with Photons,
  Leptons, and Jets}}, \href{https://doi.org/10.1007/JHEP11(2012)097,
  10.1007/JHEP07(2016)009}{\emph{JHEP} {\bfseries 11} (2012) 097}
  [\href{https://arxiv.org/abs/1204.1119}{{\ttfamily 1204.1119}}].

\bibitem{Kim:2018mks}
J.~H. Kim and I.~M. Lewis, \emph{{Loop Induced Single Top Partner Production
  and Decay at the LHC}},
  \href{https://doi.org/10.1007/JHEP05(2018)095}{\emph{JHEP} {\bfseries 05}
  (2018) 095} [\href{https://arxiv.org/abs/1803.06351}{{\ttfamily
  1803.06351}}].

\bibitem{Dawson:2012di}
S.~Dawson and E.~Furlan, \emph{{A Higgs Conundrum with Vector Fermions}},
  \href{https://doi.org/10.1103/PhysRevD.86.015021}{\emph{Phys. Rev.}
  {\bfseries D86} (2012) 015021}
  [\href{https://arxiv.org/abs/1205.4733}{{\ttfamily 1205.4733}}].

\bibitem{deBlas:2017xtg}
J.~de~Blas, J.~C. Criado, M.~P\'erez-Victoria and J.~Santiago, \emph{{Effective
  description of general extensions of the Standard Model: the complete
  tree-level dictionary}},
  \href{https://doi.org/10.1007/JHEP03(2018)109}{\emph{JHEP} {\bfseries 03}
  (2018) 109} [\href{https://arxiv.org/abs/1711.10391}{{\ttfamily
  1711.10391}}].

\bibitem{AguilarSaavedra:2009mx}
J.~A. Aguilar-Saavedra, \emph{{A Minimal set of top-Higgs anomalous
  couplings}},
  \href{https://doi.org/10.1016/j.nuclphysb.2009.06.022}{\emph{Nucl. Phys.}
  {\bfseries B821} (2009) 215}
  [\href{https://arxiv.org/abs/0904.2387}{{\ttfamily 0904.2387}}].

\bibitem{Grzadkowski:2010es}
B.~Grzadkowski, M.~Iskrzynski, M.~Misiak and J.~Rosiek, \emph{{Dimension-Six
  Terms in the Standard Model Lagrangian}},
  \href{https://doi.org/10.1007/JHEP10(2010)085}{\emph{JHEP} {\bfseries 10}
  (2010) 085} [\href{https://arxiv.org/abs/1008.4884}{{\ttfamily 1008.4884}}].

\bibitem{Aaboud:2018urx}
{\scshape ATLAS} collaboration, \emph{{Observation of Higgs boson production in
  association with a top quark pair at the LHC with the ATLAS detector}},
  \href{https://doi.org/10.1016/j.physletb.2018.07.035}{\emph{Phys. Lett.}
  {\bfseries B784} (2018) 173}
  [\href{https://arxiv.org/abs/1806.00425}{{\ttfamily 1806.00425}}].

\bibitem{Sirunyan:2018hoz}
{\scshape CMS} collaboration, \emph{{Observation of $\mathrm{t\overline{t}}$H
  production}},
  \href{https://doi.org/10.1103/PhysRevLett.120.231801}{\emph{Phys. Rev. Lett.}
  {\bfseries 120} (2018) 231801}
  [\href{https://arxiv.org/abs/1804.02610}{{\ttfamily 1804.02610}}].

\bibitem{Cepeda:2019klc}
{\scshape HL/HE WG2 group} collaboration, \emph{{Higgs Physics at the HL-LHC
  and HE-LHC}},  \href{https://arxiv.org/abs/1902.00134}{{\ttfamily
  1902.00134}}.

\bibitem{Ellis:2018gqa}
J.~Ellis, C.~W. Murphy, V.~Sanz and T.~You, \emph{{Updated Global SMEFT Fit to
  Higgs, Diboson and Electroweak Data}},
  \href{https://doi.org/10.1007/JHEP06(2018)146}{\emph{JHEP} {\bfseries 06}
  (2018) 146} [\href{https://arxiv.org/abs/1803.03252}{{\ttfamily
  1803.03252}}].

\bibitem{Atlas:2019qfx}
{\scshape ATLAS, CMS} collaboration, \emph{{Report on the Physics at the HL-LHC
  and Perspectives for the HE-LHC}},  in \emph{{HL/HE-LHC Physics Workshop:
  final jamboree Geneva, CERN, March 1, 2019}}, 2019,
  \href{https://arxiv.org/abs/1902.10229}{{\ttfamily 1902.10229}}.

\bibitem{AguilarSaavedra:2010zi}
J.~A. Aguilar-Saavedra, \emph{{Effective four-fermion operators in top physics:
  A Roadmap}}, \href{https://doi.org/10.1016/j.nuclphysb.2011.06.003,
  10.1016/j.nuclphysb.2010.10.015}{\emph{Nucl. Phys.} {\bfseries B843} (2011)
  638} [\href{https://arxiv.org/abs/1008.3562}{{\ttfamily 1008.3562}}].

\bibitem{Buckley:2015lku}
A.~Buckley, C.~Englert, J.~Ferrando, D.~J. Miller, L.~Moore, M.~Russell et~al.,
  \emph{{Constraining top quark effective theory in the LHC Run II era}},
  \href{https://doi.org/10.1007/JHEP04(2016)015}{\emph{JHEP} {\bfseries 04}
  (2016) 015} [\href{https://arxiv.org/abs/1512.03360}{{\ttfamily
  1512.03360}}].

\bibitem{Cirigliano:2016njn}
V.~Cirigliano, W.~Dekens, J.~de~Vries and E.~Mereghetti, \emph{{Is there room
  for CP violation in the top-Higgs sector?}},
  \href{https://doi.org/10.1103/PhysRevD.94.016002}{\emph{Phys. Rev.}
  {\bfseries D94} (2016) 016002}
  [\href{https://arxiv.org/abs/1603.03049}{{\ttfamily 1603.03049}}].

\bibitem{Cirigliano:2016nyn}
V.~Cirigliano, W.~Dekens, J.~de~Vries and E.~Mereghetti, \emph{{Constraining
  the top-Higgs sector of the Standard Model Effective Field Theory}},
  \href{https://doi.org/10.1103/PhysRevD.94.034031}{\emph{Phys. Rev.}
  {\bfseries D94} (2016) 034031}
  [\href{https://arxiv.org/abs/1605.04311}{{\ttfamily 1605.04311}}].

\bibitem{AguilarSaavedra:2018nen}
D.~Barducci et~al., \emph{{Interpreting top-quark LHC measurements in the
  standard-model effective field theory}},
  \href{https://arxiv.org/abs/1802.07237}{{\ttfamily 1802.07237}}.

\bibitem{Apollinari:2116337}
G.~Apollinari, I.~B\'ejar~Alonso, O.~Brüning, M.~Lamont and L.~Rossi,
  \emph{{High-Luminosity Large Hadron Collider (HL-LHC): Preliminary Design
  Report}}, CERN Yellow Reports: Monographs. CERN, Geneva, 2015,
  \href{https://doi.org/10.5170/CERN-2015-005}{10.5170/CERN-2015-005}.

\bibitem{Sirunyan:2017fho}
{\scshape CMS} collaboration, \emph{{Search for excited quarks of light and
  heavy flavor in $\gamma +$ jet final states in proton–proton collisions at
  $\sqrt{s} =$ 13TeV}},
  \href{https://doi.org/10.1016/j.physletb.2018.04.007}{\emph{Phys. Lett.}
  {\bfseries B781} (2018) 390}
  [\href{https://arxiv.org/abs/1711.04652}{{\ttfamily 1711.04652}}].

\bibitem{Bigi:1986jk}
I.~I.~Y. Bigi, Y.~L. Dokshitzer, V.~A. Khoze, J.~H. Kuhn and P.~M. Zerwas,
  \emph{{Production and Decay Properties of Ultraheavy Quarks}},
  \href{https://doi.org/10.1016/0370-2693(86)91275-X}{\emph{Phys. Lett.}
  {\bfseries B181} (1986) 157}.

\bibitem{Buchkremer:2012dn}
M.~Buchkremer and A.~Schmidt, \emph{{Long-lived heavy quarks : a review}},
  \href{https://doi.org/10.1155/2013/690254}{\emph{Adv. High Energy Phys.}
  {\bfseries 2013} (2013) 690254}
  [\href{https://arxiv.org/abs/1210.6369}{{\ttfamily 1210.6369}}].

\bibitem{Lee:2018pag}
L.~Lee, C.~Ohm, A.~Soffer and T.-T. Yu, \emph{{Collider Searches for Long-Lived
  Particles Beyond the Standard Model}},
  \href{https://doi.org/10.1016/j.ppnp.2019.02.006}{\emph{Prog. Part. Nucl.
  Phys.} {\bfseries 106} (2019) 210}
  [\href{https://arxiv.org/abs/1810.12602}{{\ttfamily 1810.12602}}].

\bibitem{Aaboud:2019trc}
{\scshape ATLAS} collaboration, \emph{{Search for heavy charged long-lived
  particles in the ATLAS detector in 36.1 fb$^{-1}$ of proton-proton collision
  data at $\sqrt{s} = 13$ TeV}},
  \href{https://doi.org/10.1103/PhysRevD.99.092007}{\emph{Phys. Rev.}
  {\bfseries D99} (2019) 092007}
  [\href{https://arxiv.org/abs/1902.01636}{{\ttfamily 1902.01636}}].

\bibitem{Aguilar-Saavedra:2013pxa}
J.~A. Aguilar-Saavedra and M.~P\'erez-Victoria, \emph{{Top couplings and top
  partners}}, \href{https://doi.org/10.1088/1742-6596/452/1/012037}{\emph{J.
  Phys. Conf. Ser.} {\bfseries 452} (2013) 012037}
  [\href{https://arxiv.org/abs/1302.5634}{{\ttfamily 1302.5634}}].

\bibitem{Aaboud:2018pii}
{\scshape ATLAS} collaboration, \emph{{Combination of the searches for
  pair-produced vector-like partners of the third-generation quarks at
  $\sqrt{s} =$ 13 TeV with the ATLAS detector}},
  \href{https://doi.org/10.1103/PhysRevLett.121.211801}{\emph{Phys. Rev. Lett.}
  {\bfseries 121} (2018) 211801}
  [\href{https://arxiv.org/abs/1808.02343}{{\ttfamily 1808.02343}}].

\bibitem{Sirunyan:2019sza}
{\scshape CMS} collaboration, \emph{{Search for pair production of vector-like
  quarks in the fully hadronic final state}},
  \href{https://arxiv.org/abs/1906.11903}{{\ttfamily 1906.11903}}.

\bibitem{Chala:2017xgc}
M.~Chala, \emph{{Direct bounds on heavy toplike quarks with standard and exotic
  decays}}, \href{https://doi.org/10.1103/PhysRevD.96.015028}{\emph{Phys. Rev.}
  {\bfseries D96} (2017) 015028}
  [\href{https://arxiv.org/abs/1705.03013}{{\ttfamily 1705.03013}}].

\bibitem{Alhazmi:2018whk}
H.~Alhazmi, J.~H. Kim, K.~Kong and I.~M. Lewis, \emph{{Shedding Light on Top
  Partner at the LHC}},
  \href{https://doi.org/10.1007/JHEP01(2019)139}{\emph{JHEP} {\bfseries 01}
  (2019) 139} [\href{https://arxiv.org/abs/1808.03649}{{\ttfamily
  1808.03649}}].

\bibitem{Branco:1986my}
G.~C. Branco and L.~Lavoura, \emph{{On the Addition of Vector Like Quarks to
  the Standard Model}},
  \href{https://doi.org/10.1016/0550-3213(86)90060-X}{\emph{Nucl. Phys.}
  {\bfseries B278} (1986) 738}.

\bibitem{AguilarSaavedra:2002kr}
J.~A. Aguilar-Saavedra, \emph{{Effects of mixing with quark singlets}},
  \href{https://doi.org/10.1103/PhysRevD.69.099901,
  10.1103/PhysRevD.67.035003}{\emph{Phys. Rev.} {\bfseries D67} (2003) 035003}
  [\href{https://arxiv.org/abs/hep-ph/0210112}{{\ttfamily hep-ph/0210112}}].

\bibitem{Atre:2011ae}
A.~Atre, G.~Azuelos, M.~Carena, T.~Han, E.~Ozcan, J.~Santiago et~al.,
  \emph{{Model-Independent Searches for New Quarks at the LHC}},
  \href{https://doi.org/10.1007/JHEP08(2011)080}{\emph{JHEP} {\bfseries 08}
  (2011) 080} [\href{https://arxiv.org/abs/1102.1987}{{\ttfamily 1102.1987}}].

\bibitem{Cacciapaglia:2011fx}
G.~Cacciapaglia, A.~Deandrea, L.~Panizzi, N.~Gaur, D.~Harada and Y.~Okada,
  \emph{{Heavy Vector-like Top Partners at the LHC and flavour constraints}},
  \href{https://doi.org/10.1007/JHEP03(2012)070}{\emph{JHEP} {\bfseries 03}
  (2012) 070} [\href{https://arxiv.org/abs/1108.6329}{{\ttfamily 1108.6329}}].

\bibitem{Barcelo:2011wu}
R.~Barcelo, A.~Carmona, M.~Chala, M.~Masip and J.~Santiago, \emph{{Single
  Vectorlike Quark Production at the LHC}},
  \href{https://doi.org/10.1016/j.nuclphysb.2011.12.012}{\emph{Nucl. Phys.}
  {\bfseries B857} (2012) 172}
  [\href{https://arxiv.org/abs/1110.5914}{{\ttfamily 1110.5914}}].

\bibitem{Cacciapaglia:2019zmj}
G.~Cacciapaglia, T.~Flacke, M.~Park and M.~Zhang, \emph{{Exotic decays of top
  partners: mind the search gap}},
  \href{https://arxiv.org/abs/1908.07524}{{\ttfamily 1908.07524}}.

\bibitem{Dobrescu:2016pda}
B.~A. Dobrescu and F.~Yu, \emph{{Exotic Signals of Vectorlike Quarks}},
  \href{https://doi.org/10.1088/1361-6471/aacbfd}{\emph{J. Phys.} {\bfseries
  G45} (2018) 08LT01} [\href{https://arxiv.org/abs/1612.01909}{{\ttfamily
  1612.01909}}].

\bibitem{Hemmick:1989ns}
T.~K. Hemmick et~al., \emph{{A Search for Anomalously Heavy Isotopes of Low $Z$
  Nuclei}}, \href{https://doi.org/10.1103/PhysRevD.41.2074}{\emph{Phys. Rev.}
  {\bfseries D41} (1990) 2074}.

\bibitem{Mohapatra:1997sc}
R.~N. Mohapatra and S.~Nussinov, \emph{{Possible manifestation of heavy stable
  colored particles in cosmology and cosmic rays}},
  \href{https://doi.org/10.1103/PhysRevD.57.1940}{\emph{Phys. Rev.} {\bfseries
  D57} (1998) 1940} [\href{https://arxiv.org/abs/hep-ph/9708497}{{\ttfamily
  hep-ph/9708497}}].

\bibitem{Mohapatra:1998nd}
R.~N. Mohapatra and V.~L. Teplitz, \emph{{Primordial nucleosynthesis constraint
  on massive, stable, strongly interacting particles}},
  \href{https://doi.org/10.1103/PhysRevLett.81.3079}{\emph{Phys. Rev. Lett.}
  {\bfseries 81} (1998) 3079}
  [\href{https://arxiv.org/abs/hep-ph/9804420}{{\ttfamily hep-ph/9804420}}].

\bibitem{Kusakabe:2009jt}
M.~Kusakabe, T.~Kajino, T.~Yoshida and G.~J. Mathews, \emph{{Effect of
  Long-lived Strongly Interacting Relic Particles on Big Bang
  Nucleosynthesis}},
  \href{https://doi.org/10.1103/PhysRevD.80.103501}{\emph{Phys. Rev.}
  {\bfseries D80} (2009) 103501}
  [\href{https://arxiv.org/abs/0906.3516}{{\ttfamily 0906.3516}}].

\bibitem{Javorsek:2001yv}
D.~Javorsek, D.~Elmore, E.~Fischbach, D.~Granger, T.~Miller, D.~Oliver et~al.,
  \emph{{New experimental limits on strongly interacting massive particles at
  the TeV scale}},
  \href{https://doi.org/10.1103/PhysRevLett.87.231804}{\emph{Phys. Rev. Lett.}
  {\bfseries 87} (2001) 231804}.

\bibitem{DeLuca:2018mzn}
V.~De~Luca, A.~Mitridate, M.~Redi, J.~Smirnov and A.~Strumia, \emph{{Colored
  Dark Matter}}, \href{https://doi.org/10.1103/PhysRevD.97.115024}{\emph{Phys.
  Rev.} {\bfseries D97} (2018) 115024}
  [\href{https://arxiv.org/abs/1801.01135}{{\ttfamily 1801.01135}}].

\end{thebibliography}\endgroup

\end{document}